\documentclass{aastex}
\usepackage{emulateapj5}

\slugcomment{ApJ in press - Revised after referee's changes}

\shorttitle{Evolution of cluster galaxies}
\shortauthors{Poggianti et al.}

\begin{document}

\title{
A comparison of the galaxy populations in the Coma and 
distant clusters: the evolution of k+a galaxies and the role of the
intracluster medium}\footnote{Based on observations made with the William Herschel Telescope 
operated on the island of La Palma by the Isaac Newton Group in the 
Spanish Observatorio del Roque de los Muchachos of the Instituto de 
Astrofisica de Canarias.}

\author{Bianca M.\ Poggianti,$^{\!}$\altaffilmark{2}
Terry J.\ Bridges,$^{\!}$\altaffilmark{3}
Y. Komiyama,$^{\!}$\altaffilmark{4}  
M. Yagi,$^{\!}$\altaffilmark{5}
Dave Carter,$^{\!}$\altaffilmark{6}
Bahram Mobasher,$^{\!}$\altaffilmark{7}
S. Okamura,$^{\!}$\altaffilmark{4} 
N. Kashikawa$^{\!}$\altaffilmark{5} 
}
\smallskip
\affil{\scriptsize 2) Osservatorio Astronomico di Padova, vicolo dell'Osservatorio 5, 35122 Padova, Italy, poggianti@pd.astro.it}
\affil{\scriptsize 3) Anglo-Australian Observatory, PO Box 296, Epping, NSW 2121, Australia}
\affil{\scriptsize 4) Department of Astronomy, University of Tokyo,
Bunkyo-ku, Tokyo 113-0033, Japan}
\affil{\scriptsize 5) National Astronomical Observatory, Mitaka, Tokyo, 181-8588 Japan}
\affil{\scriptsize 6) Liverpool John Moores University, Astrophysics Research Institute, Twelve Quays House, Egerton Wharf, Birkenhead, Wirral, CH41 1LD, UK}
\affil{\scriptsize 7) Space Telescope Science Institute, 3700 San Martin 
Drive, Baltimore MD 21218, USA \\
Affiliated with the Space Sciences Department of the European Space Agency}

\begin{abstract}
The spectroscopic properties of galaxies in the Coma cluster are
compared with those of galaxies in rich clusters at $z \sim 0.5$,
to investigate the evolution of the star formation history in clusters. 
Luminous galaxies with $M_V \leq -20$ and 
post-starburst/post-starforming (k+a) spectra which constitute a
significant fraction of galaxies in distant cluster samples are
absent in Coma, where spectacular cases of k+a spectra are found
instead at $M_V>-18.5$ and represent a significant proportion of the
cluster dwarf galaxy population. 
A simple inspection of their positions on the sky indicates that
this type of galaxy does not show a
preferential location within the cluster, but the bluest and
strongest-lined group of k+a's lies in projection towards the central
1.4 Mpc of Coma and have radial velocities significantly higher
than the cluster mean. 
We find a striking correlation between the positions of these young
and strong post-starburst galaxies and substructure in the hot
intracluster medium (ICM) identified from {\it XMM-Newton}
data, with these galaxies lying close to the 
edges of two infalling
substructures. 
This result strongly suggests that the interaction
with the dense ICM could be responsible for the quenching of the
star formation (thus creating the k+a spectrum), and possibly,  
for any previous starburst. 
The evolution with redshift of the
luminosity distribution of k+a galaxies
can be explained by a ``downsizing effect'', with the
maximum luminosity/mass of actively star-forming galaxies 
infalling onto clusters decreasing at lower redshift.
We discuss the possible physical origin of this downsizing effect
and the implications of our results for current scenarios
of environmental effects on the star formation in galaxies.

\end{abstract}


\section{Introduction}

Fundamental to our understanding of how and why galaxies evolve is a
detailed knowledge of how their stellar content evolves and what
physical processes drive these changes.  Over the years, solid
observational evidence has accumulated indicating that processes
external to galaxies can be as important as, and may be in some cases more
important than, intrinsic processes operating inside galaxies.  The
average stellar content of galaxies changes profoundly as a function
of environment, attesting that the build-up of stellar populations
occurred in a fashion that was greatly influenced by environmental
conditions.  In order to study galaxy evolution, clusters of galaxies
are therefore excellent places to start with, because their virialized
and infall regions allow us to probe a large range of physical conditions
and can be used to test the relative roles of intrinsic evolution, of
the local environment around a galaxy, and of the global environment,
on a group-, filament- and cluster-wide scale, over a wide range of
structure masses.

The evolution of galaxies in clusters has been investigated through
observations of distant clusters using galaxy colors
\citep{bo84,elli01,kb01,stan98,k98}, morphologies
\citep{dre97,f00,vd00,lubin02}, scaling relations
\citep{kelson97,kelson00,ziegler97} and spectral properties
\citep{cou87,dg92,abr96,fis98,dre99,balogh99, elli01,postman01}. In
particular, galaxy spectra carry a wealth of information about the
current and past star formation activity in galaxies and, in this
regard, are superior to photometry for studying the evolutionary
histories of galaxies.

Spectroscopic surveys of galaxies in distant clusters have detected
high fractions of currently or recently starforming galaxies at high-z
\citep{cou87,fis98,dre99,postman98,postman01}. They have shown that
star formation is suppressed as a consequence of the environment for
those galaxies that have recently accreted
\citep{dre99,pog99,abr96,balogh97}.  These studies have clarified the
clustercentric radial dependence of the spectral properties
\citep{elli01,fis98,balogh97,balogh98,balogh99,dre99}, showing the
potential of these observations to trace the infall history of
clusters and constraining the evolution with redshift of the infall
rate \citep{elli01}. Several of these studies have debated whether the
star formation activity in clusters is enhanced in some of the
infalling galaxies, before being finally quenched
\citep{dre99,balogh99,couch01}. Spectroscopic surveys of distant
clusters have elucidated the star formation histories of the various
Hubble types, underlining the lack of a one-to-one correlation between
galaxy morphology and star formation properties
\citep{dre99,pog99,couch01}.  They have compared the cluster and field
spectroscopic properties at similar redshifts
\citep{balogh98,dre99,balogh99,postman01} and those in clusters of
different X-ray luminosities \citep{balogh02}.  The first detailed
comparison of cluster galaxy spectra {\it across} a redshift range is
now being carried out \citep{tran03}.

However, none of these high-z spectroscopic studies have attempted a
{\sl quantitative} comparison with clusters in the local Universe.
So far, the evolution inferred has always been based on
the qualitatively well established result that the great majority of
galaxies in the richest clusters today are devoid of any sign of
current or recent star formation. 
The reason for this is that until now no similar study has been
carried out on a suitable local
dataset of analogous spectral quality sampling the global galaxy
populations of clusters. Surprisingly, a complete and detailed census
of the spectroscopic properties of galaxies in nearby clusters and
groups is still lacking, but should soon be provided by the
recent and ongoing surveys at low-z (e.g. Sloan, 2dF, \cite{pimbbletI,fas03}). 

In the last few years we have been carrying out a photometric and
spectroscopic study of galaxies in the Coma cluster, obtaining deep
wide-area B and R-band photometry and multifiber spectra for about 300
cluster members.  The spectroscopic survey extends over more than 6
magnitudes, including both giant and dwarf galaxies down to $M_B \sim
-14$.  This is Paper VIII of a series presenting the results of these
observations.  The photometric (Komiyama et al. 2002) and
spectroscopic (Mobasher et al. 2001) datasets were used to study the
stellar population ages and metallicities as a function of galaxy
magnitude and Hubble type (Poggianti et al. 2001a, 2001b) and of 
environment (Carter et al. 2002), to investigate the dynamics of the
dwarf and giant populations (Edwards et al. 2002), and to derive the
cluster luminosity function (Mobasher et al. 2003).

In this paper, we compare the spectroscopic properties of galaxies in
Coma with those of galaxies in clusters at $z \sim 0.5$.
The comparison of a single cluster with still relatively
small samples of clusters at higher redshift
is obviously limited. Nevertheless, we will argue that 
the gross spectral differences between the two epochs are 
more outstanding than the variance among the different
distant clusters, and that this comparison
is teaching us something about
how the star formation properties of galaxies in clusters
have evolved with redshift. 
After
a brief description of our dataset (\S2), we reanalyze our spectra in
a similar fashion to what was done for the distant cluster data (\S3).
The comparison of the spectral characteristics of luminous galaxies in
Coma and in clusters at higher redshifts
is presented in \S4.  In \S5 we extend
the spectroscopic analysis to galaxies at magnitudes fainter than can
be probed at high redshifts, and find a significant population of
post-starburst dwarfs. Their spectra and colors, incidence at
different magnitudes, radial velocities, location within the cluster
and positions relative to the substructure traced by the hot X-ray gas
are presented in \S5. In \S6 we discuss the similarities
and differences between the results presented here and previous work
on Balmer-enhanced galaxies in Coma. Finally, in \S7 we conclude with
a discussion of the possible interpretation of our results within
the current debate on environmental effects on star formation in
galaxies.

In the following we adopt $H_0 = 65 \, \rm km \, s^{-1} \, Mpc^{-1}$
and a distance modulus to Coma of 35.16.

\section{Observations}

We have carried out a spectroscopic survey of faint and bright
galaxies in two fields, each of size $32.5 \times 50.8 \, \rm
arcmin^2$, in the direction of the center of the Coma cluster (Coma1) 
where the two dominant galaxies lie (NGC4874 and NGC4889),
and in the south-west region where the cD galaxy
NGC4839 is located (Coma3).  These
areas were imaged in the B and R bands with the Japanese mosaic CCD
camera on the William Herschel Telescope (Komiyama et al. 2002, Paper
I).  A full description of the selection of the spectroscopic targets,
observations and data reduction can be found in Mobasher et al. (2001, Paper
II).  Here we only summarize the most important characteristics of the
sample and the data.

The spectra were taken with the WYFFOS multi-fibre spectrograph at the
William Herschel Telescope. They are centered on 5100 \AA $\,$ and
extend over more than $3000$ \AA $\,$ with a resolution $6-9$ \AA $\,$
FWHM. Each fibre has a diameter of 2.7 arcsec, thus
the spectra are sampling the central 1.3 kpc of each galaxy.

Targets for spectroscopy were selected from our
R-band magnitude limited sample with essentially no additional
color or morphological selection criteria, except for the exclusion of
a few (2 in Coma1 and 9 in Coma3, at most) extremely blue ($B-R<1$)
faint galaxies for which no cluster membership information is
available.
In the following we restrict our analysis to galaxies with velocities
in the range $4000<v<10000 \, \rm km \, s^{-1}$ (roughly corresponding
to $3\sigma$ cuts), which are considered members of the Coma cluster.
With this membership criterion, the spectroscopic catalog comprises a
total of 278 galaxies, of which 189 are in Coma1 and 89 in Coma3.

The survey extends over more than 6 magnitudes in galaxy luminosity, 
corresponding to a B
band absolute magnitude in the range $M_B\sim-20.5$ to $-14$ mag
for our choice of distance modulus to Coma. For convenience, in the
following we will refer to ``dwarf'' and ``giant'' galaxies
adopting a threshold $R_{3Kr}=16.3$\footnote{Magnitude over an
aperture of radius 3 times the Kron radius. Throughout the paper,
$R=R_{3Kr}$.}  corresponding approximately to $M_B=-17.3$. This
magnitude limit is expected to separate dwarf galaxies and
low-luminosity ellipticals from spiral galaxies and giant ellipticals
in most cases. Nevertheless, we stress
that this division should only be considered a way to separate the
bright and the faint subsamples and does not provide information about
any other galactic characteristic such as morphological type or
central surface brightness. According to our adopted division, the
spectroscopic catalog comprises 160 dwarf and 118 giant galaxies
members of the Coma cluster.
In the following we further restrict the sample to galaxies
brighter than $R<19$ because at magnitudes fainter than this 
the signal-to-noise of the spectra is generally too low for a
reliable spectroscopic analysis (\S3). This leaves us with a 
total of 147 dwarf and 113 giant Coma galaxies.

\section{Line measurements and classification scheme}

In previous papers of this series we have used absorption line
indices of the Lick system to estimate luminosity-weighted ages and
metallicities \citep{pog01a}. For the purposes of this paper, instead,
we classify our Coma spectra into a number of spectral
classes according to the method used for distant clusters presented in
Dressler et al. (1999, MORPHS collaboration). The MORPHS spectral
scheme allows a direct comparison between our Coma results and higher
redshift clusters, since no comprehensive dataset of Lick indices is
available yet for distant cluster galaxies.

The equivalent width of $\rm H\delta$ in absorption and of the main
emission lines (if present: [O{\sc ii}]$\lambda$3727, $\rm H\beta$,
[O{\sc iii}]$\lambda$4959,[O{\sc iii}]$\lambda$5007, $\rm H\alpha$)
were measured as ratio of the line and the
continuum fluxes, in the same way as was done for the MORPHS dataset
(Dressler et al. (1999)), after binning the spectra
at 6 \AA $\,$
to match the same rest frame wavelength resolution
of the spectra at $z=0.5$.  
The classification criteria are based only on the presence or absence
of emission lines and on the strength of
the $\rm H\delta$ line,
which is commonly in absorption.  These lines are good indicators
of current (emission lines, especially [O{\sc ii}]) and 
recent ($\rm H\delta$) star formation and
the scheme is a convenient, relatively simple method to identify the
broad star formation properties of post-starburst/post-starforming,
star-forming (spiral-like or
starbursting) and passive galaxies.  
A description of
the modeling and interpretation of each spectral class can be found in
Poggianti et al. (1999).  
In the following we will consider three spectral types:

\it k+a/a+k spectrum, \rm if no emission-line is detected 
and EW($\rm H\delta)> 3$ \AA.
\rm It is the spectrum of a post--starburst/post--starforming galaxy
with no current star formation activity which
was forming stars at a vigorous rate in the recent past (last 1.5 Gyr).

\it emission-line spectrum, \rm 
if at least one emission-line is detected. 
\rm It is the 
spectrum of a galaxy with some ongoing star formation or/and AGN activity. This
activity can be at a low, moderate or high level. In the MORPHS
scheme emission-line galaxies are further split into three different
classes, while here we adopt a simplified scheme and group them all
together.

\it k spectrum, \rm when no emission-line is 
detected and EW($\rm H\delta)< 3$ \AA. 
\rm This is the spectrum of a ``passive'' galaxy with 
neither on--going nor strong recent star formation during the last
1.5 Gyr.

We note that our spectra include either the wavelength region with
[O{\sc ii}]3727 or the region around $\rm H\alpha$. In all cases,
spectra were searched for other emission lines ($\rm H\beta$, 
[O{\sc iii}]$\lambda$4959/$\lambda$5007).  No emission line is
detected in our k+a and k spectra down to a few \AA\ EW $\,$ ($<2-5$ \AA),
which is the typical limit for detection. 
Even if some very weak emission
were present and undetected, k+a spectra do not resemble those of
normal galaxies of any Hubble type. In fact, spectra with
EW($\rm H\delta) > 3 $ \AA $\,$ and emission lines weaker than a few
Angstrom are exceptional in samples of nearby field galaxies (see
discussion in Poggianti et al. 1999, based on Kennicutt 1992, and Jansen
et al. 2000, Jansen 2002 private communication).  Moreover, normal,
non-interacting galaxies of any morphological type and {\it any emission-line
strength} typically have EW($\rm H\delta$) smaller than 4-5 \AA $\,$
(Poggianti et al. 1999). As we will show later, about half of the k+a
galaxies we identify in Coma have EW($\rm H\delta)$ even higher than 5
\AA.

\section{Comparison with distant clusters}

In this section we carry out a comparison between the spectral types of
distant cluster galaxies and those of Coma galaxies of similar
absolute magnitudes. The Coma dataset is our spectroscopic survey
described in \S2.  The distant dataset is the spectroscopic catalog
of the MORPHS collaboration (Dressler et al. 1999), that includes 10
rich clusters at z=0.4-0.5 with a wide range of properties in terms of
concentration, optical and X-ray luminosity.

The MORPHS spectroscopic catalog has a luminosity distribution
representative of cluster galaxies with an absolute V magnitude $M_V
\leq -19.9$. Therefore, in comparing with the distant clusters, we first
adopt an R-band luminosity cut in Coma ($R \sim 14.75$) corresponding
to $M_V < -19.9$.\footnote{This conversion in magnitudes assumes 
the rest-frame color of an Sbc galaxy (V-R=0.5). This is a conservative choice
because if the color of an elliptical
was adopted, the R band limit in Coma would be even brighter (14.5).}
The spectral characteristics of fainter Coma
galaxies will be discussed in the next section.
No color or morphological selection criteria were applied in the original
selection of the MORPHS galaxies. The small morphological bias (tendency for 
late-type galaxies to be slightly overrepresented) that was detected
a posteriori was corrected for in
Poggianti et al. (1999), thus the MORPHS numbers used here can be considered 
effectively drawn from a simply magnitude limited sample, and can be compared
with our Coma data.
Since on average the characteristics of galaxy spectra vary
strongly with clustercentric radius, we only consider
galaxies within a similar cluster area in all clusters, 
i.e. the central $\sim 1.4$ Mpc.

The most remarkable difference we find between the MORPHS dataset and Coma
is the fact that a large number of k+a spectra were found
in the MORPHS sample (about 20\% of the total galaxy population), while
this fraction is equal to zero (no galaxy detected) at comparable
luminosities in Coma. No k+a galaxy as luminous as those in
the MORPHS sample is detected even outside the central
1.4 Mpc of Coma, i.e. in the South-West Coma3 region with NGC 4839,
as further discussed in \S6.

To assess the influence of variation in cluster properties, 
we can limit the comparison to the two MORPHS
clusters whose X-ray luminosities more closely resemble Coma,
(Cl0016+16 and 3C 295)\footnote{The X-ray luminosity of Coma
corresponds to $L_X \sim 9.5 \, 10^{44} \, \rm ergs \, s^{-1}$ when
observed at z=0.5 at 0.3-3.5 keV and lies in between that of Cl0016+16
($L_X \sim 11.8 \, 10^{44} \, \rm ergs \, s^{-1}$, 0.3-3.5 keV) and
that of 3C 295 ($L_X \sim 6.4 \, 10^{44} \, \rm ergs \, s^{-1}$,
0.3-3.5 keV) \citep{sma97a}.}.
These two clusters have even higher k+a fractions ($\sim 30\pm10$\%) than 
the whole sample, making
the lack of bright k+a galaxies in Coma even more striking.

Neither the uncertainties in the conversion between observed and
absolute magnitudes (0.3 mag at most, 
due to the cosmological parameters and the galaxy
colors adopted), nor a plausible small luminosity
brightening in $M^{\star}$ at higher redshifts ($\sim 0.3$ mag in V in
ellipticals between z=0.55 and z=0, \citep{sma97b}) can come even
close to explaining the lack of luminous k+a's in Coma.  As we will see in
the next section, in Coma we need to go two magnitudes fainter to find
a significant k+a population.

Numerous other spectroscopic surveys have found significant populations of
k+a galaxies in distant clusters (see for example \cite{cou87,abr96,
dg92,fis98,tran03}).  We have chosen to compare with
the MORPHS sample because it is currently the largest sample of
clusters at $z>0.3$ with detailed spectroscopic analysis, and because
it is possible to adopt similar magnitude and area cuts, as well as
line measurement methods, for our Coma data.  We note however that 
much lower (but not zero) k+a fractions
than those found in the MORPHS clusters were
detected in the cluster sample of the Canadian Network for
Observational Cosmology by Balogh et al. (1999), even at redshifts
comparable to the MORPHS sample. The mean incidence of k+a's in clusters and
its variance at any cosmological epoch will be fully clarified by the
ongoing and future large surveys of high redshift clusters.
Currently, the contrast between the presence/absence of notable numbers
of luminous k+a galaxies in clusters at z=0.5/Coma is remarkable 
and points to a significant evolution with redshift of this type
of galaxy population.

Turning to emission-line galaxies, there is a tendency for these to be
less common in Coma ($9.4\pm5$\%) than in the overall MORPHS sample
($26.5\pm3$\%).  Choosing to compare only to clusters with similar
X-ray luminosities, however, strongly attenuates the differences with
Coma, since Cl0016+16 and 3C295 have emission-line fractions of
$11.1\pm6$\% and $14.7\pm8$\%, respectively.  Thus, while the k+a
incidence does not depend much on cluster properties in the MORPHS
sample, the emission-line fraction strongly does, which makes it hard to
disentangle a reliable evolutionary trend of the emission-line
fraction from a variance as a function of the cluster properties on
the basis of the two datasets we consider here.  Finally, k-type
spectra dominate the Coma cluster representing $90.6\pm17$\% in our
sample, 
compared to $\sim 51\pm$4\% in the distant cluster galaxies,
and $\sim 55\pm15$\% in Cl0016+16 and 3C 295.
Before presenting the spectroscopic properties of Coma
galaxies fainter than $M_V \sim -20$ in \S5, we first discuss 
the possible effects of the spectroscopic aperture on our
spectral classifications and the implications for the conclusions
reached so far.

\subsection{Radial color gradients in galaxies and aperture effects}

A reason for concern when comparing nearby and distant clusters is the
difference in the galactic area (in kpc) covered by the spectra at low
and high redshift. The fibre diameter of our Coma spectra corresponds
to the central 1.3 kpc of the galaxies, while the slit spectrum at
z=0.5 samples $\sim 7 \times 18$ kpc.  Therefore, in principle, if the
central regions of galaxies were older than the outer regions, the
higher fraction of passive (k-type) galaxies in Coma compared to
distant clusters and the lack of k+a spectra
could be due to aperture effects.

The star formation gradients observed in cluster galaxies actually go
in the opposite direction.  A number of studies have found evidence that
the most recent episode of star formation tends to be 
{\sl more centrally concentrated}
in cluster galaxies than in field samples,
both in star-forming and post-starforming galaxies
\citep{moss93,koo98,moss00,rose01,bartholomew01,koo03}. 
If this is generally the case, aperture effects would tend to 
{\sl underestimate} the evolution with redshift.

However, we prefer to evaluate the impact of aperture effects on a
galaxy-by-galaxy basis from our own data, employing radial color
gradients within individual 
galaxies in Coma. Here we use the fact that color and
spectral type are largely correlated and that if recent or ongoing star
formation occurred in the outer regions of the galaxy (those not
encircled within the fibre), these would tend to be bluer than the
galaxy centre. 

The B-R colors over an aperture equal to the fibre diameter (2.7
arcsec) are plotted against the total (3 Kron radii) colors in Fig.~1.
For the comparison with the distant clusters, only the k-type galaxies
with $R<14.75$ (left panel, filled dots) are relevant.  All these
luminous k-type galaxies have colors $(B-R)_{2.7''}>1.4$, and since in
this case color and spectral type are derived from the same aperture, a
color redder than 1.4 can be considered to correspond to a k-type
spectrum. If the outer regions of a galaxy were significantly younger
and displayed a more ``active'' spectral type, then its total color
should be significantly bluer than 1.4, by more than 0.1 mag, as
expected from the color-magnitude diagram in Fig.~2 and the colors of
galaxies with various spectral types.  This is obvious for the k
versus emission-line types, but it is a solid argument also for
k+a's. In fact, as can be seen also in Fig.~2, k+a galaxies can have a
wide range in color, from blue ``young'' k+a's (observed soon after
the termination of star formation) to ``old'' ones as red as
ellipticals (after about 1-1.5 Gyr from truncation) \citep{cou87}.
Although a lack of a color gradient therefore does not guarantee the
absence of a red, old k+a stellar population in the outskirts of the
galaxy, if some of the bright k-classified galaxies had k+a stellar
populations outside of the fibre, statistically some of them (the
young ones) should show a bluing at increasing distance from the
galaxy center.  We note that metallicity gradients will alter the
color gradients as well, but their effects will be much more subtle
than the large color differences implied by the coarse spectral typing
adopted here.

Thus, a galaxy with an old centre and young outskirts
should lie in Fig.~1 at $(B-R)_{tot}<1.4$ {\it and} $(B-R)_{2.7''}>1.4$;
however, no significant population with these
characteristics is observed, with at most one or two bright 
galaxies showing evidence for a gradient of this kind.
This reinforces the conclusion based on age gradients from other studies
that aperture effects are not likely to be responsible for the gross
differences in spectral properties between Coma and the high redshift
sample. We conclude that the observed spectral differences 
are real and that Coma has
proportionally many more passive galaxies than the distant clusters,
while lacking a significant population of luminous k+a's.

\section{Exploring lower luminosities}

Any comparison with the distant cluster data is inevitably limited to
the bright end of the galaxy luminosity function, while in nearby
clusters it is possible to study also fainter galaxies.  In Coma, our
spectroscopic survey extends over more than 6 magnitudes down to $M_B
\sim -14$. In the following, we limit our discussion to $R<19$ ($M_B <
-14.6$) because at magnitudes fainter than this limit a large fraction
of the galaxies have an uncertain spectral type due to the low
signal-to-noise of the spectra.

In the previous section we showed that Coma lacks the population
of {\sl luminous} k+a galaxies that is present in clusters at
z=0.4. Interestingly, very clear examples of k+a spectra are found
instead among the {\sl faint} Coma galaxies.

Figure~3 presents the spectra of all our ``secure'' k+a galaxies, with
good $\rm H\delta$ measurements confirmed by the strength of other
Balmer lines. The mean signal-to-noise of these spectra is 11.4/pixel,
ranging from a maximum of 17/pixel (\#90085) to a minimum of 6/pixel
(\#48397)\footnote{The S/N has been measured as an average of the blue
and red sides of the $\rm H\delta$ line.}.

The k+a spectra can be contrasted with a k-type spectrum in Fig.~4.
We have chosen to show the k spectrum of a single galaxy with a
magnitude similar to the k+a spectra, instead of coadding several
spectra into a ``representative k-type'' spectrum, because in this way
the S/N is comparable to that of the k+a galaxies in Fig.~3.  The
difference in $\rm H\delta$ and other Balmer line strengths between the
k spectrum and the k+a's is striking even from a simple visual
inspection of the spectra. This is quantified by the measurements of the
EW($\rm H\delta$) in Table~1. The table lists the galaxy identification
number in our catalog, identification number in the catalog of
\cite{gmp83}, coordinates, R
magnitude, B-R color and $\rm H\delta$ strength of the 13 secure k+a
galaxies in our sample and of other 10 more uncertain but likely
k+a's, these latter being marked with a colon (``:'') in column 6. The
k+a spectra cover a range of $\rm H\delta$ strengths, from very
strong, text-book cases of post-starburst galaxies such as \#28211,
\#30153 and \#31086 to weaker cases with EW($\rm H\delta$) just above
3 \AA $\,$, such as \#90085 and \#47098, whose line is just strong
enough to belong to the k+a class. 

Magnitude distributions and color-magnitude diagrams of k, k+a and
emission-line galaxies are presented in Fig.~2. Two aspects of this
plot are remarkable.  First, Coma k+a's in our sample are typically
fainter than $M_V \sim -18.6$, or $R \sim 16$, i.e. more than about 3
magnitudes fainter than $M^{\star}$.
The majority of them are classified as ``dwarfs'' according to our
adopted definitions ($R > 16.3$), with the exception of two and three
``giant'' galaxies in the Coma3 and Coma1 fields, respectively.  As
pointed out in \S2, the adopted dwarfs/giants division is somewhat
arbitrary and is expected to only roughly separate dwarf galaxies and
low-luminosity ellipticals from spiral galaxies and giant ellipticals.
In the Coma1 field, the magnitude distribution of k+a's seems to
cluster around a ``characteristic luminosity'' ($M_V \sim -18/-17.5$)
(Fig.~2), but it is hard to establish whether the lack of very faint
k+a's is simply due to increasingly poor S/N spectra. The deficit of
k+a's among the brightest galaxies is instead a firm result.

Second,
in the color-magnitude diagram of Fig.~2 a group of blue and a group
of red k+a's can be easily distinguished. These most likely
correspond to ``young'' k+a's (observed soon after the termination of
star formation) and to ``old'' ones (observed at a later stage of the
evolution). In agreement with this interpretation and with results of
models of k+a spectra (e.g. \cite{cou87,pog96}), the average EW($\rm
H\delta$) of the blue group is significantly stronger than that of the
red group (Table~1). 

An equivalent width EW($\rm H\delta)>5$ \AA $\,$ testifies that a starburst
occurred in the galaxy before star formation was quenched, as a simple
interruption of otherwise quiescent star formation activity
produces a spectrum with weaker Balmer lines
(e.g. \cite{cou87,pog96}). Most of the blue k+a's, and a few of the
red k+a's, have EWs stronger than 5 \AA $\,$, and thus must truly be
post-starburst galaxies.

Morphological classifications based on visual inspection of deep
images are available from \cite{bei03} for a subset of our k+a sample:
these are mostly disky galaxies with spiral or merger/peculiar (M/P)
morphologies, two of them being classified as M/P, one as an Sc, two
as Sa's and one as S0/a. Only one red k+a has an available
classification from \cite{bei03} and it is an S0 galaxy.  
A visual inspection of our images (Komiyama et al. 2002) confirms
that k+a's are an heterogeneous population spanning the whole range of
morphological types from early to late types and mergers.
K+a spectra are found in galaxies of all Hubble types 
also in distant clusters, where the majority of k+a's are disky galaxies 
(Wirth et al. 1994, Couch et al. 1998, Dressler et al. 1999, Poggianti et
al. 1999). However, we should remember that our k+a's are fainter than
those found in the distant clusters and that most of them are dwarfs to 
which a classification early/late does not apply straightforwardly.   
In our images, most of the k+a's are found to have companions
and/or signs of tidal interactions, and most companions are much
smaller objects than the already small k+a's.
It is an open question whether these small companions
have some effect on, or are somehow related to the k+a phenomenon. 
An automated assessment of the light profiles all our k+a galaxies
is provided by the measurements of the Sersic index $N$ presented
in Paper I. With the exception of two blue k+a's compatible with
a de Vaucouleurs's profile ($N=0.25$), all the other blue and red
k+a's have exponential or steeper profiles ($N \ge 0.5$), as the 
great majority of all Coma galaxies of similar luminosities do.
A more detailed analysis of k+a and Coma dwarf morphologies in 
general will be the subject of a future work."

Estimates of luminosity-weighted metallicities are available for k+a
galaxies from our earlier analysis based on spectral indices of the
Lick system. In Paper III we found a large metallicity
scatter among galaxies at any given luminosity fainter than R=16. This
is true also for both blue and red k+a's, that have metallicities from
above solar to [M/H]$\sim -1.5$, and also for k-type galaxies of similar
luminosity. Hence, the k+a galaxies do not stand out for their
metal content with respect to the rest of the population of galaxies
of similar luminosity, and share the broad range in metallicity
of all faint Coma galaxies.

We now examine the overall incidence of the various spectral classes
in the dwarf and giant subsamples separately.\footnote{The
spectroscopic dwarf and giant subsamples have different completeness
rates (Mobasher et al. 2001), but {\sl within} each subsample this
rate is essentially flat with magnitude. Furthermore,
the magnitude range and the relative sampling as a function of galaxy
magnitude are similar for the Coma1 and Coma3 spectroscopy.
Hence, the relative proportions of
spectral types in each subsample can be compared.}  Numbers and
fractions of galaxies of each spectral type are given in Tables~2 and
3 for the Coma1, Coma3 and Coma1+Coma3 regions.
Table~3 shows that
k+a's constitute a significant fraction (10-15 \%) of the dwarf galaxy
population at $M_B > -17.3$, being much more common among the
``dwarfs'' than among the ``giants''.  If we only include galaxies
with secure spectral classifications, the k+a fraction among the
dwarfs remains similar (16$\pm$5\%).  
It is of course unknown whether a k+a
population of dwarf galaxies exists also at higher redshift, and if
they are more, equally or less frequent than at z=0.

\subsection{K+a dwarfs and substructure}

In their spectroscopic survey of early-type galaxies in Coma,
\cite{cal93} detected enhanced Balmer lines in absorption in a large
fraction of the galaxies in the South-West region of Coma, mostly
between the cluster centre and the secondary peak of the X-ray
emission in the SW, in a region that has not been extensively sampled
in our survey.  They found instead very few such galaxies in the
central field of the cluster. A spatial segregation of galaxies with
peculiar star formation signatures can obviously be useful in
identifying substructure, and can serve to investigate the effects of
cluster-group merging on the properties of galaxies.

In Figure~5 we show the projected position on the sky of galaxies of
different spectral types in our sample. From a simple inspection of this
figure, k+a galaxies do not seem preferentially located in any of the
regions we surveyed.  As shown in Table~3, the k+a fractions are
similar within the errors in Coma1 and Coma3. However, those with the
strongest $\rm H\delta$ EWs (the blue k+a's) are all situated in
projection towards the Coma1 central field, though only one of them is
in the central projected 25 arcmin from the two central
galaxies. We will show later in this section that blue k+a's are not
randomly distributed in the cluster, and that a strong correlation 
with substructure becomes evident from a deeper analysis.

Interestingly, the mean radial velocity of the group of blue 
k+a's is 8120$\pm 709$ km $\rm s^{-1}$, with all but one at $v>7200$
km $\rm s^{-1}$ (see Table~1).  In contrast, both the red k+a's and
all faint k galaxies with $R>16$ in Coma1 have much lower mean
velocities: 6992$\pm 761$ and 6854$\pm 244$ km $\rm s^{-1}$,
respectively\footnote{The difference between the mean velocities of
k+a's and all faint k's in Coma3 is small, 6999$\pm 859$ versus
7134$\pm 319$ km $\rm s^{-1}$. Most k+a's in Coma3 are red
(Fig.~2). Mean velocities and dispersions have been computed using the
biweight estimator computed with the program ROSTAT by \citep{beers90}
(see also Teague et al. 1990).}.  The hypothesis that the group of
blue k+a's and the other two groups are drawn from the same parent
velocity distribution is rejected by a Kolmogorov-Smirnov test at the
93.3\% and 99.995\%, respectively.

The strength of the lines and the color of the blue k+a's indicate
a halting of the star formation within the last 500 Myr and probably
at some point during the last $\sim 5 \times 10^7$ to $3 \times 10^8$
yr. In most cases (those
with EW($\rm H\delta)>5 \AA$) a strong starburst must have preceded
the quenching of star formation \citep{pog99,pog96}. In contrast, the
typical time elapsed since the last star formation in the red k+a's
will be in the range 700-1500 Myr.  The fact that the youngest k+a's are a
kinematically distinct population from the general relaxed dwarf
population (k-types) and older k+a's suggests that their star
formation history was recently quenched as a result of their
first interaction with the main cluster body.  

The blue k+a galaxies have a relatively high velocity dispersion (1250
$\rm km \, s^{-1}$) indicating that probably
they are not part of a single infalling bound
galaxy group of relatively small mass compared to the whole Coma
cluster. 
Their velocities and velocity dispersion 
suggest instead they could be part of a filament converging onto Coma.

Proof of this, and 
a suggestive clue about the possible physical mechanism responsible
for the k+a spectra, comes from the recent X-ray mosaic observations of
Coma obtained with {\it XMM-Newton}. \cite{neu03} have recently
identified and discussed X-ray substructure by fitting a smooth
profile (an elliptical beta model of a relaxed cluster) and
subtracting it from the data. The residuals reveal several structures,
that are shown as contours in Fig.~6: besides the well known NGC4839
South-West group, \cite{neu03} identify a large residual to the West
of the cluster centre (``Western structure'' in Fig.~6) elongated
along the North-South direction, and a filament-like structure
South-East of the centre (``Eastern structure'' in Fig.~6), elongated
along the East-West direction.  The temperature map shown in color in
Fig.~6 sheds further light on the accretion history of
Coma. Neumann et al. conclude that the region of high temperature
observed between the Western structure and the Coma center is caused
by the infall of this structure, either via compression or via shock
waves.  These authors consider the two maxima in the western structure
to be likely the result of the disruption of a galaxy group during its
infall, instead of two galaxy groups falling at the same time.  In
contrast, the South-Eastern structure is cooler than the mean cluster
temperature and is associated with a low-mass galaxy group
dominated by two large galaxies, NGC4911 and NGC4921.  Based on the
filamentary form of this structure, the same authors conclude it is
observed during the infall process while being affected by ram
pressure stripping close to the cluster centre.

The coincidence of the position of the strongest k+a galaxies and the
X-ray structures is striking. Four k+a's with EW($\rm H\delta)>5$ \AA
$\,$ (green dots in Fig.~6) trace the edge of the Western structure
towards the Coma centre. Another three are associated with the
Eastern structure, all at its western boundary.  One k+a with
uncertain classification (\#6750, k+a: in Table~1) is at the location
of a strong temperature gradient between a hot and a cool region, though
\cite{neu03} warn that such hot spot could be a statistical artefact
due to the low {\it XMM-Newton} exposure time in that region.

The correlation between the location of the post-starburst galaxies
and the substructure in the intracluster medium appears too strong to
be a fortuitous coincidence. Young post-starbursts are distributed
close to the edge of infalling substructures. In the case of the Western
substructure this edge is the infalling front, while for the Eastern
substructure it is unclear whether the group is moving to the West,
as suggested by the appearance of the X-ray residuals, or
to the East, as suggested by the positions of NGC4911 and NGC4921
\citep{neu03}.  Overall, this strongly suggests that the k+a spectra,
i.e. the truncation of the star formation activity in these galaxies
and possibly the previous starburst, could be the result of an
interaction with the hot intracluster medium.
Hence, as far as Coma k+a galaxies are concerned, there is suggestive evidence
that the origin of the k+a spectrum is a cluster-related and, in 
particular, an ICM-related phenomenon that is closely connected
with the dynamical state of the cluster.
We note that the $\rm H\delta$ strength of 
the blue k+a galaxies implies that star
formation was truncated in these galaxies 
on a short timescale, i.e. short compared
to the k+a timescale of 1-1.5 Gyr. In fact, a slowly
declining star formation activity such as that envisaged if
galaxies simply lost their gas halo reservoir when becoming
part of a group (``strangulation'', e.g. Bower \& Balogh 2003)
does not produce such strong Balmer lines.

The two large galaxies that dominate the Eastern
structure (NGC4911 and NGC4921) have velocities of 7973 and 7560 km/sec,
respectively, hence they too have velocities significantly higher than
the cluster median, as do the blue/strong k+a's. This is consistent with
the hypothesis that the three strong k+a's close to the Eastern
structure are physically associated with the NGC4911/4921 group. It
remains to be interpreted why the other strong/blue k+a's, and
especially those four along the Western structure, also have similar
velocities, and whether the two structures are somehow associated
and/or related to the NGC4839 group and on larger scales with the
filament connecting Coma and A1367 along the Great Wall.

We stress here the importance of isolating the {\it youngest}
k+a galaxies using their blue colors and strong equivalent widths. The
red k+a phase has a timescale that is comparable to the core crossing
time in a cluster like Coma, and any signature of the link between the
truncation of star formation and the location within a substructure is
thus erased in the older k+a's, while it is still detectable in the
youngest subsample of k+a's.

Obviously, establishing a correlation between the star formation
history of the k+a galaxies and the substructure within Coma would
have been impossible on the basis of simple location on the sky
(Fig.~5), while this correlation appears evident once a detailed X-ray
map reveals the complicated structure in the hot intracluster gas.  On
the other hand, by isolating the strongest/youngest post-starburst
galaxies we have identified specific areas in the cluster where a
physical mechanism must have been recently at work affecting the star
formation activity of these galaxies.

\subsubsection{Emission-line galaxies}

Interestingly, the few emission-line galaxies in the Coma1
field (all starbursts or star-forming galaxies of unknown star
formation intensity) are mostly found along the edge of the Western
structure (Fig.~7), with the exception of \#15480 that is located just South
of the Eastern structure.  However, only two of the Coma1 emission-line
galaxies lie at high velocities (7525 and 9443 $\rm km \, s^{-1}$,
galaxies \#15480 and \#77938, respectively), while the other four have
velocities between 5007 and 5666 km/sec. The two high
velocity emission-line galaxies (both starbursting galaxies)
could be associated with the same
subsystems to which the majority of the blue k+a galaxies belong.
The other Coma1 emission-line galaxies remain a puzzle:
their alignment with the X-ray substructure suggests they too could be related
to the same infalling group, but their low velocities seem to contradict this
hypothesis.

Emission-line galaxies are proportionally more numerous
in the Coma3 than in the Coma1 field (Figs.~5 and ~6).
For dwarfs we estimate a 13\% fraction of emission-line galaxies in
Coma3 versus 5\% in
Coma1, and among giants 33\% in Coma3 versus 9\% in Coma1 (Table~3). It
should be noted that the emission-line properties of dwarf and giant
spectra are on average very different, because the former mostly
present moderate or strong lines, while the latter tend to have very
weak emission, sometimes barely detectable. Hence, we stress that the
higher emission-line fractions among giants should not be considered
as indicative of a generally higher star formation activity in
brighter galaxies.

\subsection{Faint k+a galaxies and aperture effects}

Similarly to what has been done in \S4.1, we estimate the relevance of
aperture effects in order to assess the reliability of the comparison
between giants and dwarfs. More luminous galaxies are on average
bigger in size and the fibre is sampling proportionally a smaller
portion of the galaxy. Could the observed higher proportion of k+a's
among dwarfs be caused by aperture effects? For this to be the case,
the k+a stellar populations should be preferentially located outside
of the central region of the galaxy.
As discussed also in \S4.1, the latest star formation episode in
cluster galaxies, and post-starburst/post-starforming galaxies in
particular, is found to be instead centrally concentrated
(\cite{rose01,bartholomew01}, see also \cite{norton01} for field
k+a's). Moreover, color gradients can again be used to evaluate stellar
population gradients within galaxies.  The left panel of Fig.~1
presents the central versus total color of all k-type galaxies, both
brighter and fainter than $R=14.75$. The arguments discussed in \S4.1
apply here as well: we do not observe a significant population of
galaxies whose color gradients suggest a noticeable difference between
the spectral type in the region covered by the fibre and the
integrated spectral type, within the gross spectral classification
scheme adopted here. Only very few galaxies have a total color that is
significantly bluer than the central color and which corresponds to
star-forming or blue post-starforming galaxies. Galaxies with a
central k-type and an outer k+a {\sl red} population at a late stage
of its evolution would go unnoticed in this analysis. Thus, in
principle we cannot exclude the existence of giant galaxies with red
k+a spectra in their outer regions that went undetected in our data.

Central versus total colors of k+a galaxies are shown in the right
panel of Fig.~1. As discussed above, these tend to be generally faint
galaxies, hence the fibre spectrum covers a significant fraction of
the galactic area. Most of these galaxies have a negligible color
difference between the central 2.7'' and the outskirts, while the 
two galaxies with larger color gradients are bluer
in the center.

\section{Comparison with previous work on Coma and other nearby clusters}

Though a low fraction of k+a spectra among
luminous galaxies in nearby clusters is often
cited \citep{dre87}, a quantitative comparison of the incidence of
this type of galaxy in nearby and in distant clusters has not been
carried out to date, due to the lack of large samples of galaxy
spectra in low redshift clusters suitable for this purpose.
This is the first time that such a comparison is attempted.

Spectra with enhanced Balmer lines in absorption have been previously reported
in early-type galaxies in 
Coma and other nearby clusters in a number of papers by Caldwell,
Rose and collaborators (e.g. Rose et al. 2001 and references therein).
Out of the 12 galaxies in the sample of Balmer-enhanced galaxies
without emission studied by \cite{cal93} (hereafter C93), six are in
common with our Coma spectroscopic survey. The comparison of the
results is rather instructive and shows that our k+a galaxies
and their Balmer-enhanced galaxies have on average significantly
different Balmer line intensities. 
In fact, only 1 out of the 6 galaxies in common
qualifies to be classified as k+a in our spectra\footnote{This
is galaxy D216, \#90085 in this work. The C93 multifiber
spectra cover an area only slightly smaller than ours (2''), and 
they use the CN/H8 index, which is a fit to the slope of the spectrum
between 3858 and 3893 \AA, to identify Balmer-enhanced galaxies.}, 
while the other five galaxies have EW($\rm H\delta)<3$ \AA $\,$ in our data.
However, 4 of these galaxies (D21, D146, D127, D14), although with a
k-type spectrum, do show signs of recent star formation based on the
analysis of age- and metallicity-sensitive line indices of the Lick
system measured in our spectra (Poggianti et al. 2001a), in agreement
with Caldwell et al.'s detection of recent activity.\footnote{There is
only one case (galaxy D34) where our and Caldwell et al. 1993 results
disagree because we do not detect any sign of star formation during
the last 9 Gyrs, neither with the MORPHS method nor using the Lick
system.}. In these galaxies the latest star formation episode must
have been weaker or is older than in galaxies with a k+a spectrum, but
it can still be detected by a spectral analysis that is sensitive to even
weaker/older activity, such as that performed by Caldwell and
collaborators and by us in Poggianti et al. (2001a).\footnote{This is
true also when comparing with the sample of Caldwell et al. (1998),
where both their and our 2001 work identify galaxies with young
(G2603), intermediate-age (G3205, G3126) and old (G2385, G3707)
stellar populations, with an excellent agreement on the
luminosity-weighted ages derived by the two studies. None of these
galaxies shows a k+a spectrum in our data.}  This is in a sense
expected since the Caldwell et al. 
sample is composed {\sl only of morphologically selected 
early-type galaxies}, while ours is a magnitude limited sample
with no color or morphological selection criteria and, as discussed
in \S5, the strongest k+a galaxies turn out to have late-type
and merger/peculiar morphologies. Their sample
is therefore probably
more skewed towards galaxies at a later stage in their path to
become more passive/earlier-type galaxies. In fact, \cite{cal93} note
that these galaxies are similar to the {\sl red} $\rm H\delta$ strong
galaxies in distant clusters, i.e. those observed at a late stage of
their Balmer-strong phase. The lower line/burst strength in 
their Balmer-enhanced galaxies 
as compared to distant clusters was also pointed out by \cite{cal97}
for Coma and other four nearby clusters.
It is interesting to note that \cite{cal93} also stressed that
the majority of their Balmer-enhanced galaxies are in a 
luminosity range unexplored in distant clusters because they are too faint.

A main result from \cite{cal93} was that the Balmer-enhanced galaxies
were mostly located in a region South-West of the center of Coma,
where the incidence of such galaxies was much higher than in the
central region of the cluster. 
However, {\sl even fainter} galaxies with recent star
formation were not found preferentially in the SW region, but spread
all over, also to the NE of the central galaxy \citep{cal98}.
The same is true for the work of Castander et al. (2001), who found
five post-starburst galaxies out of 196 Coma members
with no apparent correlation with the NGC 4839 group.
 
To summarize, while the k+a galaxies we discuss in this paper are
examples of Balmer-strong galaxies above a certain line-strength
threshold and below a certain ``age'', Balmer-enhanced galaxies amply
discussed in previous papers include also weaker/older cases and, on
average, have weaker lines than k+a's. We suspect that these differences
in spectral properties are related to the different selection strategy for 
our sample (which had no color or morphological selection criteria)
compared to previous samples, which were limited to early-type
morphologies. We also remind the reader that 
different spectral classification schemes were adopted: the EW of
$\rm H\delta$ in our study versus the CN/H8 index (a fit to 
the slope of the spectrum between 3858 and 3893 \AA) for Balmer-enhanced
galaxies in \cite{cal93}. 
However, both our k+a's and the Balmer-enhanced spectra identified by others
are found in galaxies fainter than the k+a's observed at high redshift. 
The regions sampled in this work and in \cite{cal93} only partly
overlap in the South-West region of the cluster, thus we are unable to
confirm 
the excess of Balmer-enhanced spectra these authors
found in that specific area. 
Within the regions we have sampled, 
k+a galaxies apparently do not show a preferential spatial location in 
projection on the sky, but 
the group of youngest/strongest k+a's 
lie in projection towards the {\it central region} 
of Coma, though they avoid the cluster centre, and, as discussed in \S5,
their positions are closely related to X-ray substructure.
In addition, they stand out for
their high recession velocities compared to the cluster mean.
Finally, this is the first time an attempt is made to quantify the
incidence of k+a spectra in the galaxy populations of Coma as a whole
and as a function of galaxy luminosity, on a sample of galaxies selected
irrespectively of morphology. Thus, no direct comparison regarding
this can be made with other work.

\section{Discussion}

The results presented in this paper raise one compelling question:
why there are luminous k+a's in clusters at high redshift while these are
notably absent in Coma, where only fainter k+a's are observed?

Given the magnitude limits of high-z spectroscopic surveys, 
it is unknown how many {\it faint} k+a's exist
in clusters at z=0.5, as the luminous ones could simply be the tip of
the iceberg.  Trying to infer this indirectly from the spectroscopic
ages of our Coma dwarfs is difficult due to the uncertainties
involved, especially those regarding the evolution of the infall rate
of galaxies onto Coma, the spectral type mix of the infalling
population and the timescale for the halting of star formation
once a star-forming galaxy enters a cluster.

In principle, as some of the physical mechanisms affecting galaxies in
dense environments are expected to act differently depending on galaxy
mass, the cause of the star formation truncation in luminous k+a's at
high-z and in faint ones at low-z is not necessarily the same.
However, those physical properties of clusters that are most relevant
for their influence on galaxies (such as the effects of encounters with
other cluster galaxies and the properties of the intracluster medium) 
have not evolved
significantly between z=0.5 and z=0. Thus, environmental effects on
galaxies {\it of a given mass} should be similar at the two epochs.
It is therefore unlikely that changes in cluster physical conditions
can explain the evolution in the {\it luminous} k+a population. One
cluster property that is expected to vary with redshift is the average
amount of substructure, which should increase at higher
redshift. Shocks and/or changes of the tidal gravitational
field during group-cluster and cluster-cluster merging are
expected to significantly affect star formation in galaxies
(e.g. Bekki 1999).
However, if the change with redshift in the average amount of substructure 
were the {\it cause of the evolution} of the luminous k+a population,
then luminous k+a's should also be found in merging
clusters at low-z.  There isn't any low redshift cluster to date with
a reported high k+a fraction among luminous galaxies. Large samples
and a systematic investigation of the spectroscopic properties of
galaxies in nearby clusters will further clarify this issue (Fasano et
al. 2003).

More probably, the evolution in the incidence of luminous k+a's in
clusters is due to an evolution with redshift in the properties
of galaxies infalling into clusters, and
reflects a cosmic ``downsizing effect'', for which some
evidence has emerged in the last years: going to lower redshift,
the maximum luminosity/mass
of galaxies with significant star formation activity seem to decrease,
possibly both in clusters (Smail et al. 1998, Bower
et al. 1999, Kodama \& Bower 2001, Poggianti et al. 2001a) and in the
field (Cowie et al. 1996, Kauffmann et al. 2003).  In fact, if star
formation at higher z was active on average in more massive galaxies
than at lower z, regardless of environment, and star formation is
quenched as a consequence of the infall onto massive and dense
environments, then the k+a phenomenon should be conspicuous in
luminous galaxies in distant clusters and insubstantial
among luminous galaxies in nearby clusters, as is observed. 

Even more intriguing is searching for the physical cause of the
downsizing effect.  There can be several possible reasons for the fact
that star formation in low mass galaxies seems to be more protracted
on average than in massive galaxies.  This could simply be due to an
evolution ``intrinsic'' to galaxies. It has been suggested, for
example, that the star formation efficiency could increase with dark
matter halo mass, perhaps as a result of supernova feedback processes,
resulting in the majority of massive galaxies not forming stars at the
present day (Kauffmann et al. 2003).  Star formation in faint galaxies
is also expected to be inhibited by the UV background radiation 
after reionization ($z<7-12$) and before the background dropped ($z < 1$),
preventing rapid cooling in this period
\citep{bab92,ben02}. This effect becomes progressively more important
towards fainter galaxies, and can thus be responsible for a slower or
suppressed early star formation in the dwarfs \citep{ski03},
leaving a gas reservoir for later star formation.
In this case, the cause of the downsizing effect
would be external to galaxies (i.e. the background radiation field),
but not necessarily related to the environment.
 
Or could the downsizing effect itself be an ``environmental'' effect?
Recent studies have found a change in the star formation properties of
galaxies at a relatively low critical threshold in local galaxy
density, interpreting this result as evidence that suppression of star
formation occurs in groups as a consequence of the environment (Kodama
et al. 2001, Lewis et al. 2002, Gomez et al. 2003).  Going to lower
redshift, proportionally more galaxies infalling onto a rich cluster
like Coma are expected to have resided and therefore to possibly have
been ``pre-processed'' (=have their star formation activity switched
off) in moderately massive environments such as groups or small
clusters.  If the growth of mass structure in the Universe had an
important role in establishing the observed decline in star formation
rate from z=2 to the present (Bower \& Balogh 2003, and references
therein), it could be responsible for a decline in the number of
luminous star-forming galaxies available and, as a consequence, for
the lack of luminous k+a's at low z.  However, if this effect worked
independently of the galaxy mass, then we shouldn't be observing k+a's
in Coma at all, even at low luminosities, and thus shouldn't observe
a down-sizing effect. In fact, why would the
star-forming progenitors of the faint k+a's not have been
quenched/pre-processed as more luminous galaxies were?  
Less massive
galaxies are expected to be actually {\it more strongly} disrupted by
environmental influences than more massive galaxies, due to their
smaller gravitational potential.

If the growth of structure had a significant impact on the evolution
of the global cosmic star formation, in order for this scenario to be
simultaneously responsible for the evolution of the luminous k+a
population {\it and} consistent with the downsizing effect, the
environmental conditions acting on galaxies should vary with the
galaxy mass.  This could occur, for example, if on average dwarf
galaxies infalling into clusters today have inhabited in the past
significantly different (less massive) environments than infalling
massive galaxies did.  It will be interesting to
investigate the history of merging halos using high-resolution
simulations and explore how this depends on present-day galaxy mass.

Observing k+a galaxies in Coma we are probably
witnessing the transformation of late-type star-forming or
starbursting dwarfs into dwarf spheroidals/ellipticals as a
consequence of the cluster environment. Evidence for such a
transformation from blue star-forming to red passive dwarfs in
clusters comes also from a comparison of the faint end of
the luminosity function at z=0.2 and z=0 \citep{wilson97}. The blue k+a's
are a case in which we directly observe this transformation
process still ongoing in a nearby cluster.

Another puzzling question remains open at this point: why are most of
the blue k+a's we observe in Coma truly post-starburst systems, as
testified by the exceptionally strong Balmer lines? What triggered
the starburst? Again, the nature
versus nurture options lie open: was the burst an episode related to
processes intrinsic to the galaxy itself, or was it induced by
external, environmental effects? An intrinsic cause would mean that
many of the unperturbed dwarfs unaffected by the cluster environment
should be in a starburst phase, i.e. have experienced a discontinuous
star formation history with recent periods of enhanced activity.
Spectroscopic surveys of dwarfs in lower density environments should
be able to provide the answer.
If environment was instead the cause of the starburst,
detailed studies of the group of blue
k+a's (their morphologies, immediate surroundings, distribution of
star formation activity and of neutral and molecular gas within the
galaxy, stellar and gaseous kinematics etc.) might help in understanding 
what physical mechanism triggered the burst and, in particular, if 
it was the interaction with the intracluster medium as discussed in \S5.

To conclude, it is possible that the kind of ``activity'' observed in
distant clusters (k+a spectra, or recently terminated star formation)
is still ongoing today as it was in the past, but is simply affecting
galaxies with a different range of masses. This could reflect a change
with redshift of the typical mass of a star-forming galaxy infalling
onto a rich cluster.  The evolution of this typical mass could be
driven either by intrinsic or by ``environmental'' effects, i.e.
by the previous ``environments'' experienced by the infalling galaxies  
throughout their evolution.

\section{Summary}

In this paper we have presented a quantitative comparison
between the spectroscopic properties of galaxies in Coma and those in
rich clusters at z=0.5.  The main results can be summarized as
follows:

a) Coma lacks the population of luminous ($M_V \leq -20$) galaxies with
post-starburst/post-starforming (k+a) spectra that is present in most
distant cluster samples.

b) In contrast, a numerous population of k+a galaxies is present in
Coma at fainter magnitudes, representing a significant fraction of the
cluster dwarfs. These faint k+a's clearly divide into a red and a blue
group, probably in an advanced and an early stage of their k+a phase,
respectively. The blue k+a's are also those with the strongest Balmer
lines in absorption and the great majority of them have recently undergone
a strong starburst. These blue k+a's display spiral or merger/peculiar
morphologies. 

c) While the red k+a's are found both in the central region and in a
region South-West of the centre, the blue k+a's are all found in
projection towards the central 1.4 Mpc. Generally, k+a's
are not observed towards the cluster centre.  
The radial velocities of the blue/strong k+a's are much higher
than the cluster mean, indicating they are a kinematically distinct
population from the general relaxed dwarf population. Their high
velocity dispersion (1250 $\rm km \, s^{-1}$) argues against them
being part of a single virialized infalling group and already
suggests they could belong to a substructure converging onto Coma.
A comparison with the X-ray residuals and temperature map shows
that the young, strong post-starburst galaxies are located at the
edges of two substructures, suggesting that interaction
with the intracluster medium is the most likely cause of
the interruption of the star formation and, possibly, of the 
previous starburst.

d) We have discussed our results in the context of previous work on
Balmer-enhanced galaxies in Coma. The k+a galaxies discussed here have
on average stronger Balmer lines in absorption than Balmer-enhanced
galaxies, and probably represent the youngest/strongest examples of a
common phenomenon, and the closest local counterparts to the distant
k+a population.

e) The different incidence of luminous k+a galaxies at z=0.5 and z=0
likely reflects an evolution in the properties of the infalling
galaxies instead of a change in the physical mechanisms at work in
clusters.  We suggest that the lack of luminous k+a's in Coma and the
presence of faint ones provides further evidence of a ``downsizing
effect'': the maximum luminosity/mass of galaxies with significant star
formation activity seems to decrease at lower redshifts.  The
physical origin of this downsizing effect is unknown.  It could be
driven by the intrinsic evolution of galaxies, or could be related to
an ``environmental effect'' connected with the growth of structure in
the Universe and the progressively more massive environments inhabited
by galaxies as evolution proceeds.  This latter is a viable scenario
only if less massive galaxies infalling into clusters today inhabited
in the past, on average, less dense/massive environments than infalling
massive galaxies did.

\acknowledgments 
We thank Andrea Biviano, Richard Bower, Erica
Ellingson, Jim Rose, Ian Smail, Jacqueline van Gorkom
and the anonymous referee
for useful discussions and/or suggestions that helped 
improving this paper.

\begin{table}
\begin{center}
\caption{List of k+a galaxies\tablenotemark{a}.\label{tbl-2}}
\begin{tabular}{clcccccll}
\tableline\tableline
ID  & $\rm ID_{GMP}$ & RA(J2000) & DEC(J2000) & R & $(B-R)_{total}$ & vel. & type & EW($\rm H\delta$)  \\
\tableline
&&&&&&&& \\
\multispan{9}{\hfil{Blue k+a galaxies in Coma1}\hfil}\cr
&&&&&&& \\
45473 & 3911 & 12 58 55.4 & 27 53 14.58 & 18.79 & 1.25 & 7955 & k+a  &9.9$\pm$1.6 \\
28211 & 3176 & 12 59 46.3 & 27 44 46.13 & 17.26 & 1.09 & 9739 & k+a  &9.4$\pm$0.5  \\
30153 & 4060 & 12 58 42.5 & 27 45 37.85 & 16.46 & 0.99 & 8711 & k+a  &9.1$\pm$0.6  \\
90411 & 3640 & 12 59 15.1 & 28 15 02.96 & 16.60 & 1.00 & 7343 & k+a  &8.5$\pm$3.5  \\
31086 & 2923 & 13 00 08.2 & 27 46 23.42 & 16.57 & 1.04 & 8664 & k+a  &7.4$\pm$0.4  \\
6750  & 2498 & 13 00 43.9 & 27 35 36.72 & 17.57 & 1.27 & 7244 & k+a: &6.3$\pm$3.8 \\
29362 & 3071 & 12 59 56.2 & 27 44 47.64 & 16.22 & 1.14 & 8931 & k+a  &6.0$\pm$0.9  \\
90085 & 2989 & 13 00 02.9 & 28 14 24.94 & 16.05 & 1.23 & 7670 & k+a  &3.3$\pm$0.3  \\
58548 & 3146 & 12 59 48.5 & 27 58 57.80 & 17.67 & 1.32 & 5312 & k+a  &3.3$\pm$0.7 \\
&&&&&&&& \\
\multispan{9}{\hfil{Red k+a galaxies in Coma1}\hfil}\cr
&&&&&&&& \\
63244 & 4003 & 12 58 48.2 & 28 01 08.4 & 17.03 & 1.64 & 7074 & k+a: &5.3$\pm$2.0 \\
31300 & 3750 & 12 59 06.7 & 27 46 19.2 & 16.44 & 1.51 & 6292 & k+a: &3.8$\pm$1.1 \\
67536 & 2511 & 13 00 43.0 & 28 03 14.4 & 18.21 & 1.69 & 6783 & k+a  &3.7$\pm$0.4  \\
33963 & 3092 & 12 59 55.0 & 27 47 45.6 & 16.04 & 1.57 & 8016 & k+a::&3.3$\pm$5.0 \\
47098 & 2736 & 13 00 21.6 & 27 53 52.8 & 16.74 & 1.49 & 4856 & k+a  &3.2$\pm$0.3 \\
52689 & 2692 & 13 00 24.7 & 27 55 37.2 & 16.78 & 1.62 & 7966 & k+a: &3.1$\pm$0.8 \\
78794 & 3129 & 12 59 50.2 & 28 08 38.4 & 16.58 & 1.54 & 6849 & k+a::& ---\tablenotemark{d}    \\
&&&&&&&& \\
\multispan{9}{\hfil{k+a galaxies in Coma3}\hfil}\cr
&&&&&&&& \\
48397  & 4183 & 12 58 34.1 & 26 54 00.0 & 18.17 & 1.37 & 8114 & k+a	 & 6.9$\pm$2.9 \\ 
149036 & 4215 & 12 58 31.7 & 27 23 42.0 & 17.48 & 1.29 & 7737 & k+a\tablenotemark{b} & 6.5$\pm$1.3 \\
169748 & 4597 & 12 57 54.5 & 27 29 27.6 & 15.02 & 1.36 & 4932 & k+a:\tablenotemark{b} & 4.8$\pm$2.5\\
76219  & 5065 & 12 57 09.1 & 27 01 48.0 & 18.87 & 1.12 & 7024 & k+a:	 & 4.2$\pm$1.2\\
18086  & 4453 & 12 58 10.1 & 26 46 37.2 & 17.77 & 1.47 & 6748 & k+a    & 3.7$\pm$0.4\\ 
154595 & 5250 & 12 56 47.8 & 27 25 15.6 & 15.64 & 1.40 & 7703 & k+a:	 & 3.0$\pm$1.0\\
101924 & 5068 & 12 57 08.6 & 27 10 22.8 & 17.81 & 1.45 & 6005 & k+a::	 & ---\tablenotemark{d} \\
&&&&&&&& \\
\tableline
\tableline
\end{tabular}
\tablenotetext{a}{Possible very weak $\rm H\alpha$ emission at the
1 \AA $\,$ level.}
\tablenotetext{b}{Other 7 weakest cases (\#55373, 176486, 66948, 95392, 38741,
57916, 81862 in Mobasher et al. (2001)) are not listed here.}
\tablenotetext{d}{The $\rm H\delta$ line is unmeasurable, but other Balmer lines suggest it could be a k+a.}
\end{center}
\end{table}

\begin{table*}
{\scriptsize
\begin{center}
\centerline{\sc Table 2}
\vspace{0.1cm}
\centerline{\sc Numbers of spectra of Coma galaxies}
\vspace{0.3cm}
\begin{tabular}{lccccc}
\hline\hline
\noalign{\smallskip}
 {Cluster} & N$_{\rm tot}$ & k & k+a/a+k & emiss & ? \cr
\hline
\noalign{\smallskip}
Dwarfs 1        &  ~97     & ~26(68) &   ~8(13)  & ~2(4) &  ~12(12) \cr
Dwarfs 3        &  ~50     & ~19(34) &   ~2(5)   & ~5(6) &  ~5(5) \cr
\noalign{\smallskip}				
 Total dwarfs   &   ~147   & ~45(102)&  ~10(18)  & ~7(10)& ~17(17) \cr
\hline						
\noalign{\smallskip}				
Giants 1        & ~80      & ~63(70) &  ~2(3)    & ~4(7) &  ~0(0) \cr
Giants 3        &  ~33     & ~19(20) &  ~1(2)    & ~4(11)&  ~0(0) \cr
\noalign{\smallskip}				
 Total giants   &   ~113   & ~82(90) &   ~3(5)   & ~8(18)& ~0(0) \cr
\hline						
 \noalign{\smallskip}				
 Total          &  ~260    &~127(192)& ~13(23)   &~15(28)&  ~17(17) \cr
\noalign{\smallskip}
\noalign{\hrule}
\noalign{\smallskip}
\multispan{6}{\hfil{``1'' and ``3'' indicate the numbers in 
the Coma1 and Coma3 regions.}\hfil}\cr
\noalign{\smallskip}
\multispan{6}{\hfil{Only galaxies with $R<19$ are included.}\hfil}\cr
\noalign{\smallskip}
\multispan{6}{\hfil{? spectra have a poor quality and
no spectral class could be assigned.}\hfil}\cr
\noalign{\smallskip}
\multispan{6}{\hfil{Numbers without/within brackets exclude/include
the uncertain (:) spectra of each class.}\hfil}\cr
\noalign{\smallskip}
\end{tabular}
\end{center}
}
\vspace*{-0.8cm}
\end{table*}

\begin{table*}
{\scriptsize
\begin{center}
\centerline{\sc Table 3}
\vspace{0.1cm}
\centerline{\sc Fractions of galaxies as a function of the spectral type}
\vspace{0.3cm}
\begin{tabular}{lccccc}
\hline\hline
\noalign{\smallskip}
 {Cluster} & N$_{\rm tot}$ & k & k+a/a+k & emiss & (?) \cr
\hline
\noalign{\smallskip}
Dwarfs 1        &  ~85 &   0.80$\pm$0.10 &   0.15$\pm$0.04 &  0.05$\pm$0.02 & 0.12 \cr
Dwarfs 3        &  ~45 &   0.76$\pm$0.13 &   0.11$\pm$0.05 &  0.13$\pm$0.05 & 0.10 \cr
\noalign{\smallskip}					     
 Total dwarfs   & ~130 &   0.78$\pm$0.08 &   0.14$\pm$0.03 &  0.08$\pm$0.02 & 0.12 \cr
\hline							     
\noalign{\smallskip}					     
Giants 1        & ~80  &   0.88$\pm$0.10 &   0.04$\pm$0.02 &  0.09$\pm$0.03 & 0.00 \cr
Giants 3        &  ~33 &   0.61$\pm$0.14 &   0.06$\pm$0.04 &  0.33$\pm$0.10 & 0.00 \cr
\noalign{\smallskip}					     
 Total giants   & ~113 &   0.80$\pm$0.08 &   0.04$\pm$0.02 &  0.16$\pm$0.04 & 0.00 \cr
\hline							     
 \noalign{\smallskip}					     
 Total          & ~243 &   0.79$\pm$0.06 &   0.09$\pm$0.02 &  0.12$\pm$0.02 & 0.07 \cr
\noalign{\smallskip}
\noalign{\hrule}
\noalign{\smallskip}
\multispan{6}{\hfil{``1'' and ``3'' indicate the numbers in 
the Coma1 and Coma3 regions.}\hfil}\cr
\noalign{\smallskip}
\multispan{6}{\hfil{Fractions were calculated excluding ? and 
including : spectra.}\hfil}\cr
\noalign{\smallskip}
\end{tabular}
\end{center}
}
\vspace*{-0.8cm}
\end{table*}

\begin{figure}
\plotone{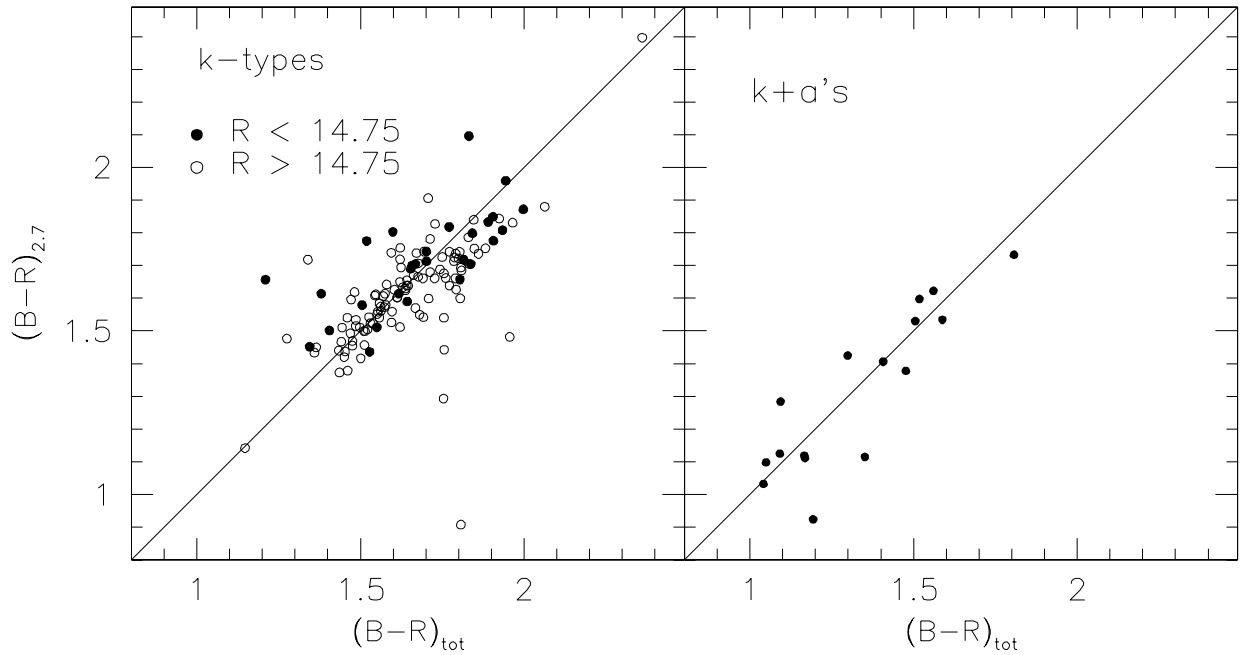}
\caption{Central (2.7'' diameter aperture)
versus total colors of k galaxies (left) and k+a's
(right). Only Coma1 galaxies are presented because
Coma3 photometry suffered from saturation in the
central pixels of the brightest galaxies.
For clarity, errorbars on the colors 
are not shown. These are typically 0.05 mag.
\label{fig3}}
\end{figure}

\begin{figure}
\plotone{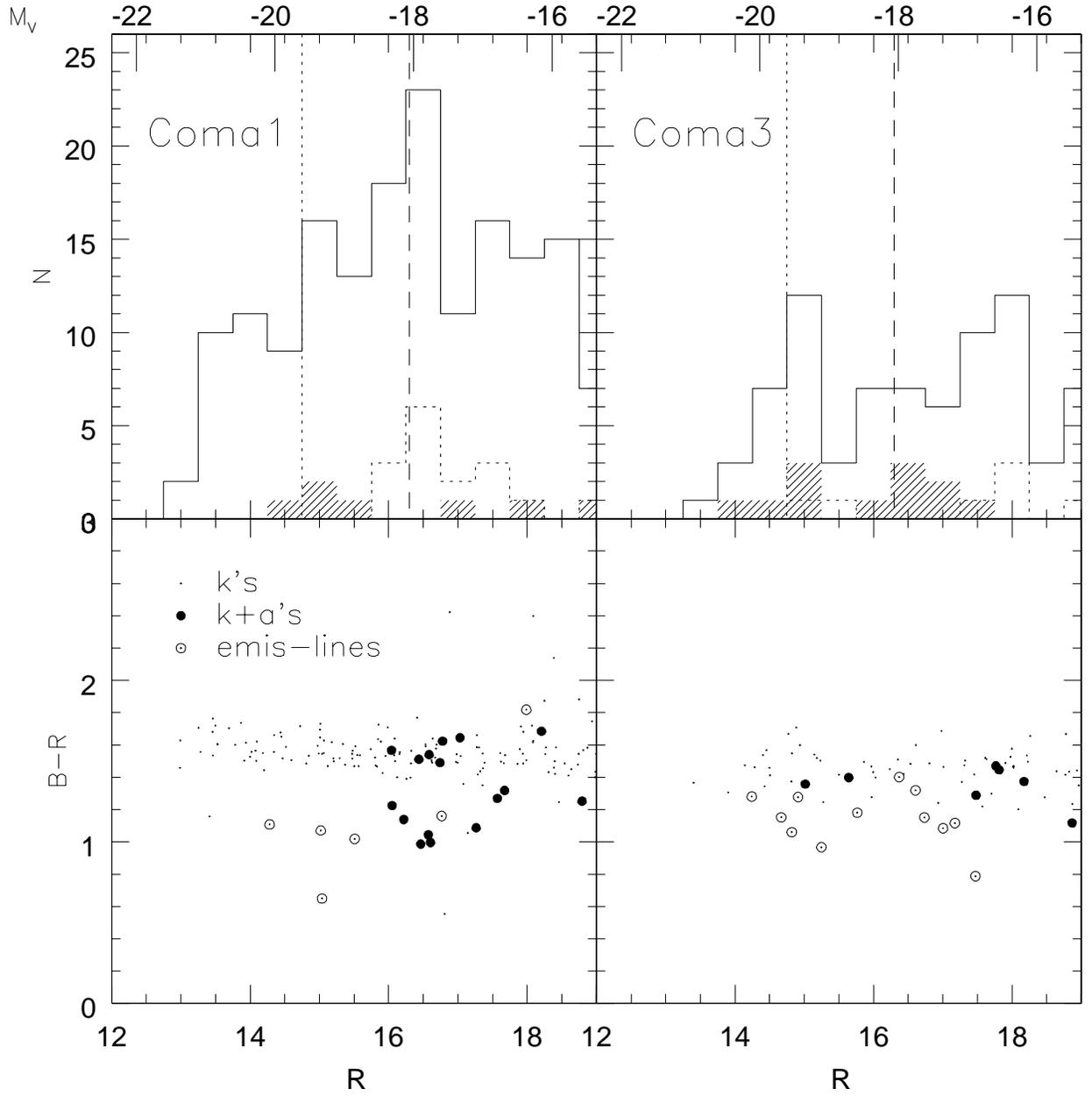}
\caption{Magnitude distributions (top) and color-magnitude diagrams
(bottom) of passive, k+a and emission-line galaxies with a securely detected
line (EW([O{\sc{ii}}]$>2$ \AA).
The dotted line
shows the magnitude limit corresponding to the MORPHS cut. The dashed
line is our adopted magnitude division between ``giants'' and ``dwarfs''.
Total histogram= all galaxies with an assigned spectral type. 
Dashed histogram = k+a's. Shaded histogram =
emission-line galaxies.
\label{fig2}}
\end{figure}

\clearpage

\begin{figure}
\centering
\includegraphics[width=0.40\columnwidth,angle=0,clip]{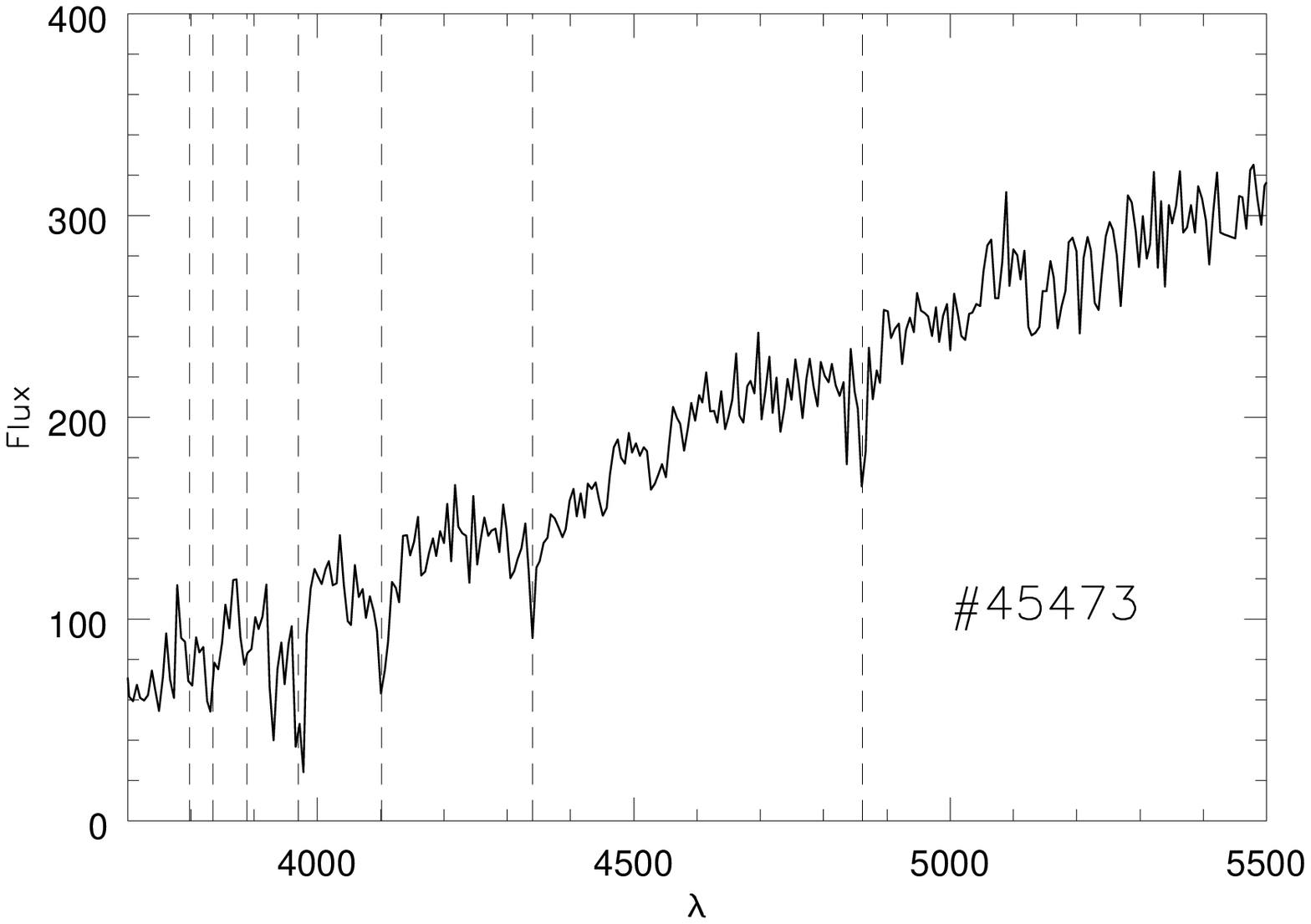}
\includegraphics[width=0.40\columnwidth,angle=0,clip]{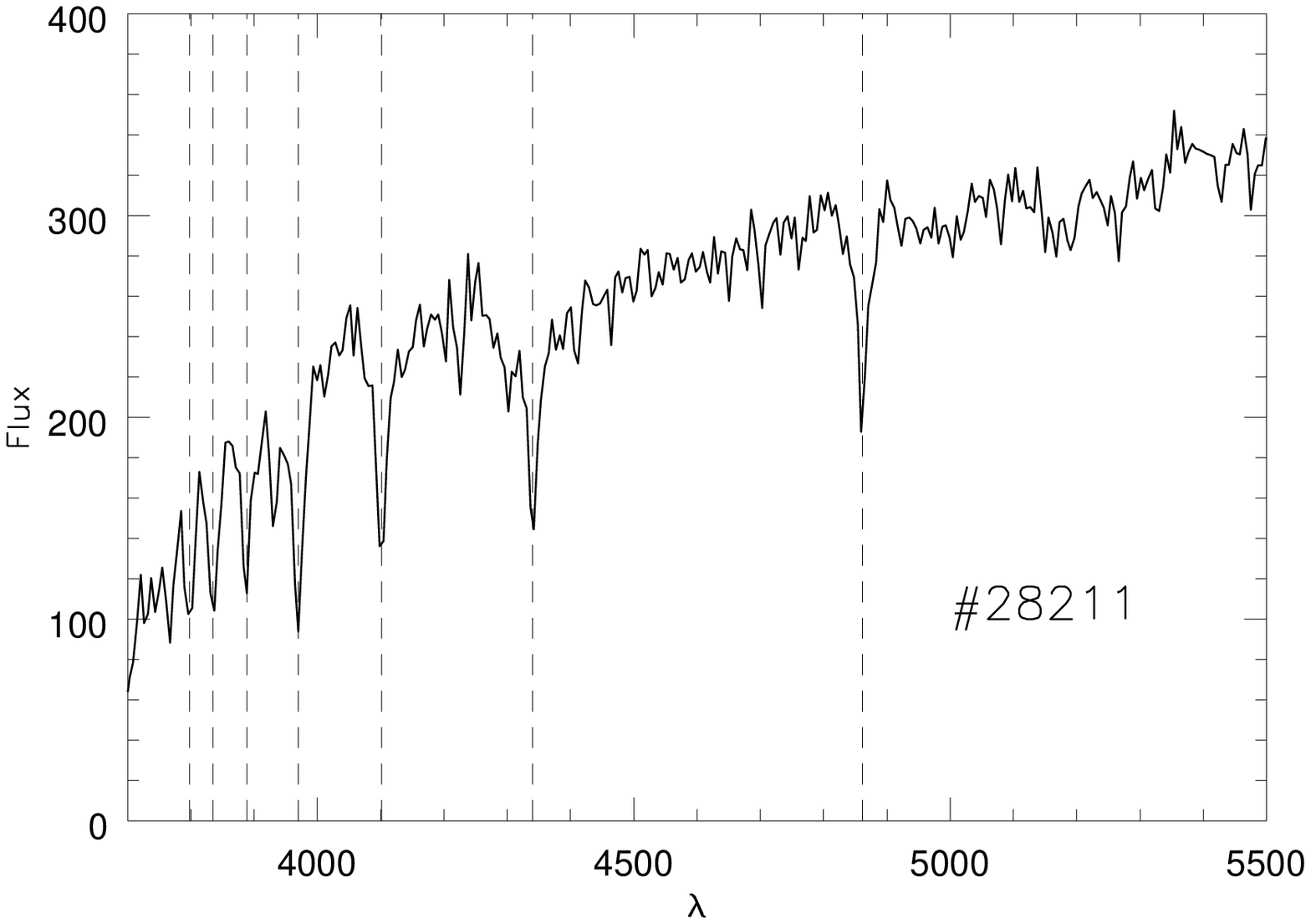}
\includegraphics[width=0.40\columnwidth,angle=0,clip]{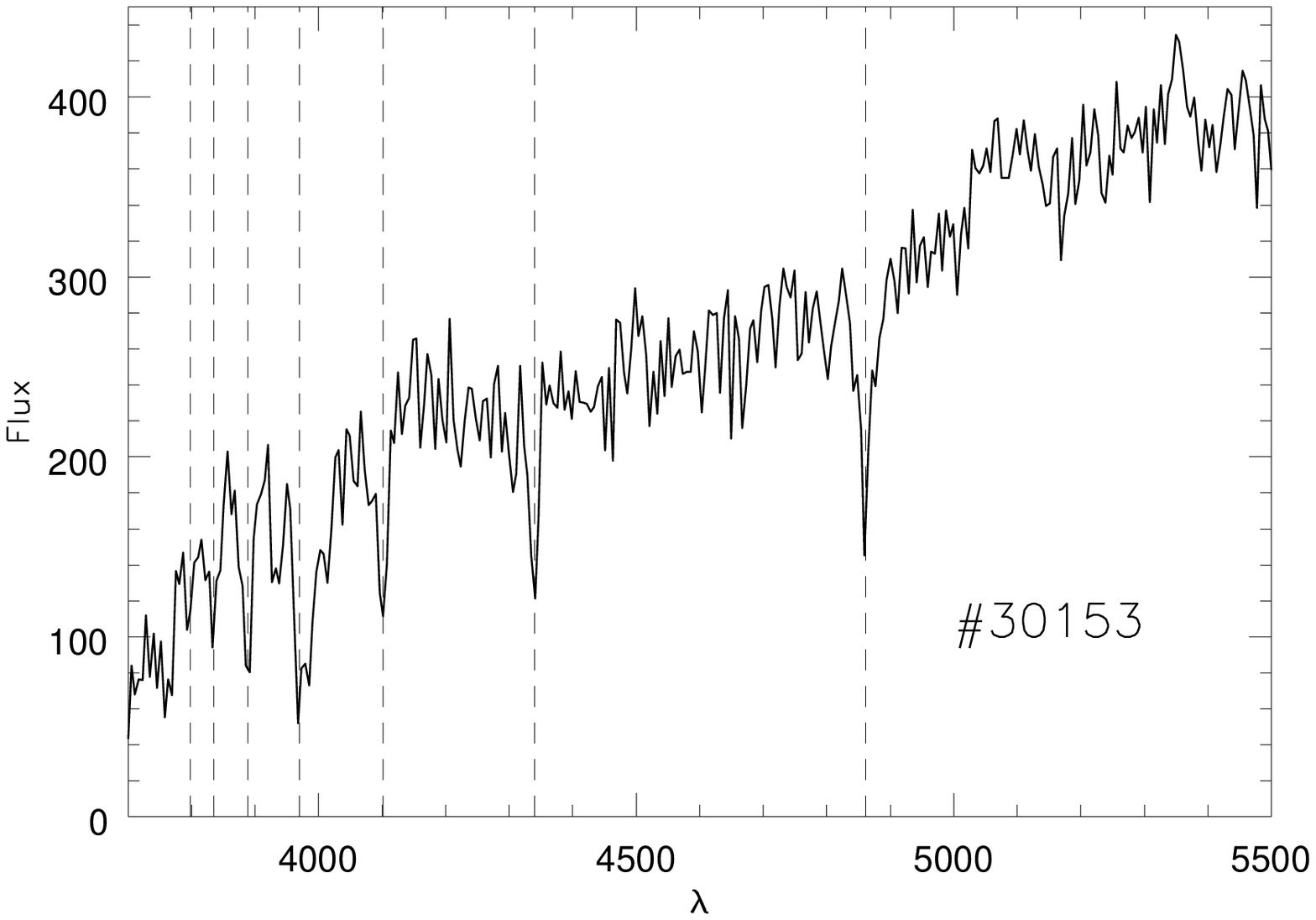}
\includegraphics[width=0.40\columnwidth,angle=0,clip]{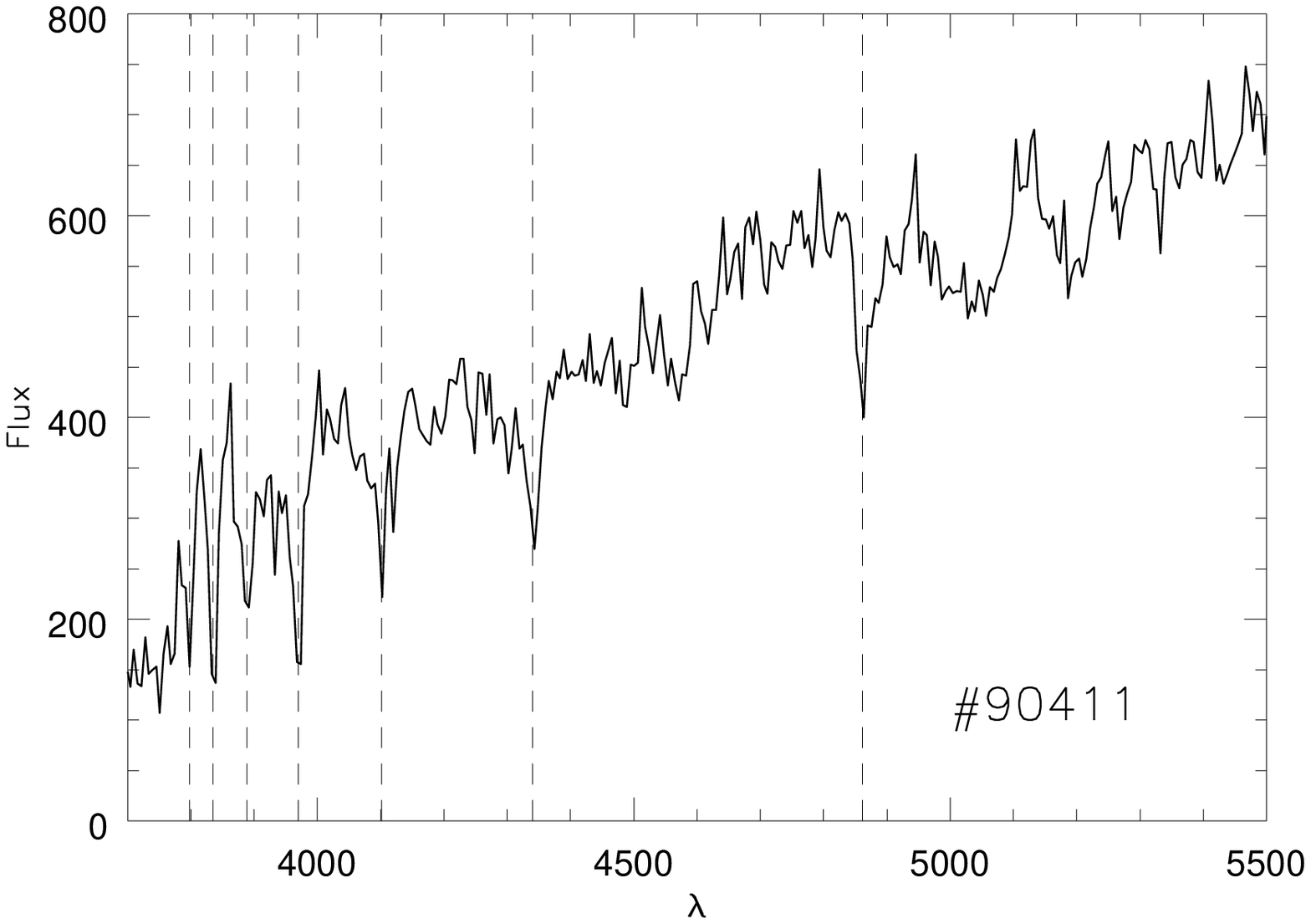}
\vskip 0pt
\caption{Rest-frame spectra of k+a galaxies in Coma rebinned at 6 \AA. 
The Balmer lines (from left
$\rm H\theta$, $\rm H\eta$, $\rm H\zeta$, $\rm H\epsilon$, $\rm H\delta$,
$\rm H\gamma$, $\rm H\beta$) are highlighted by dashed lines. 
In some cases the spectrum includes $\rm H\alpha$, but
the section redder than 5500 \AA $\,$ is not displayed.
\label{fig1}}
\end{figure}

\begin{figure}
\centering
\includegraphics[width=0.40\columnwidth,angle=0,clip]{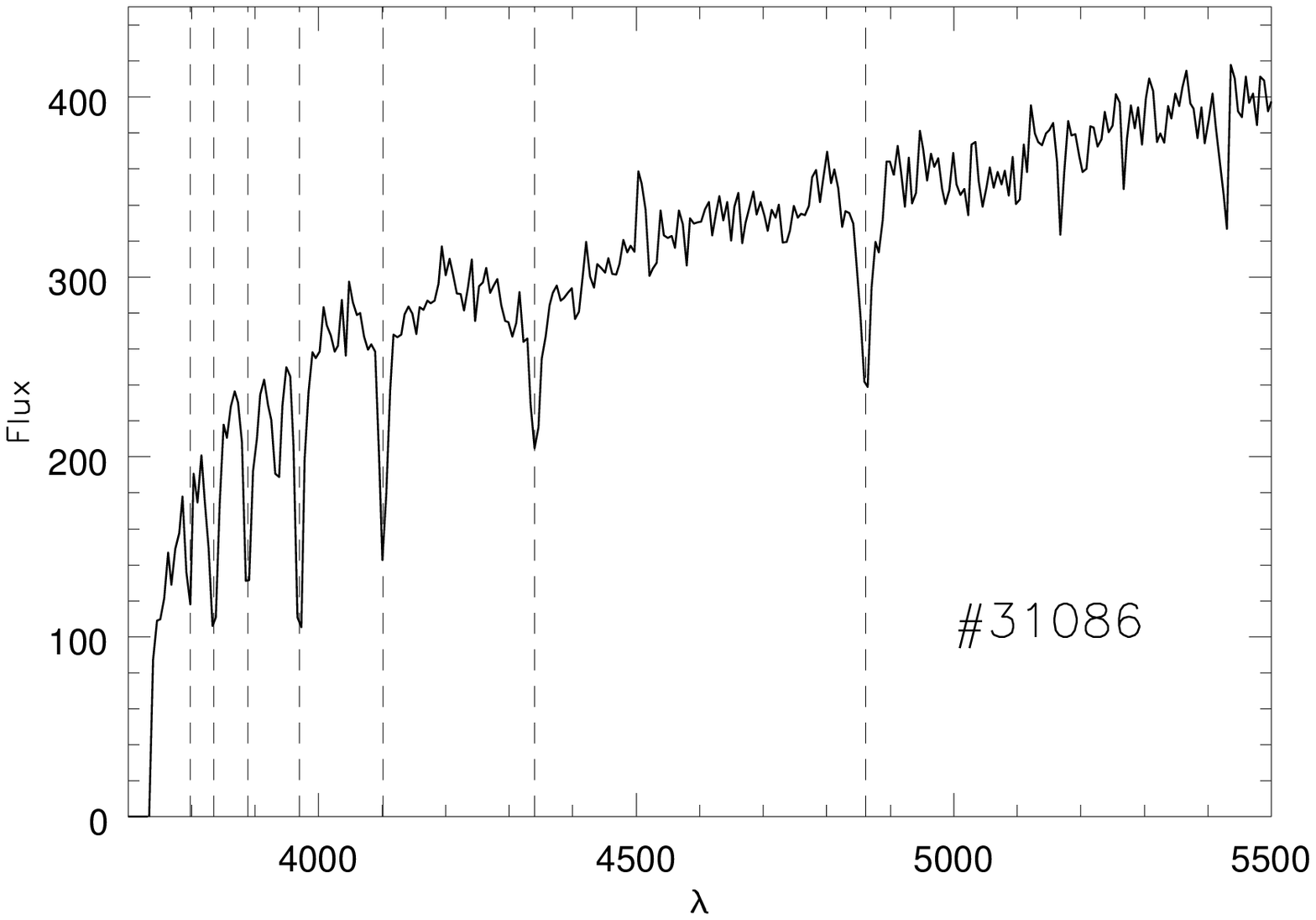}
\includegraphics[width=0.40\columnwidth,angle=0,clip]{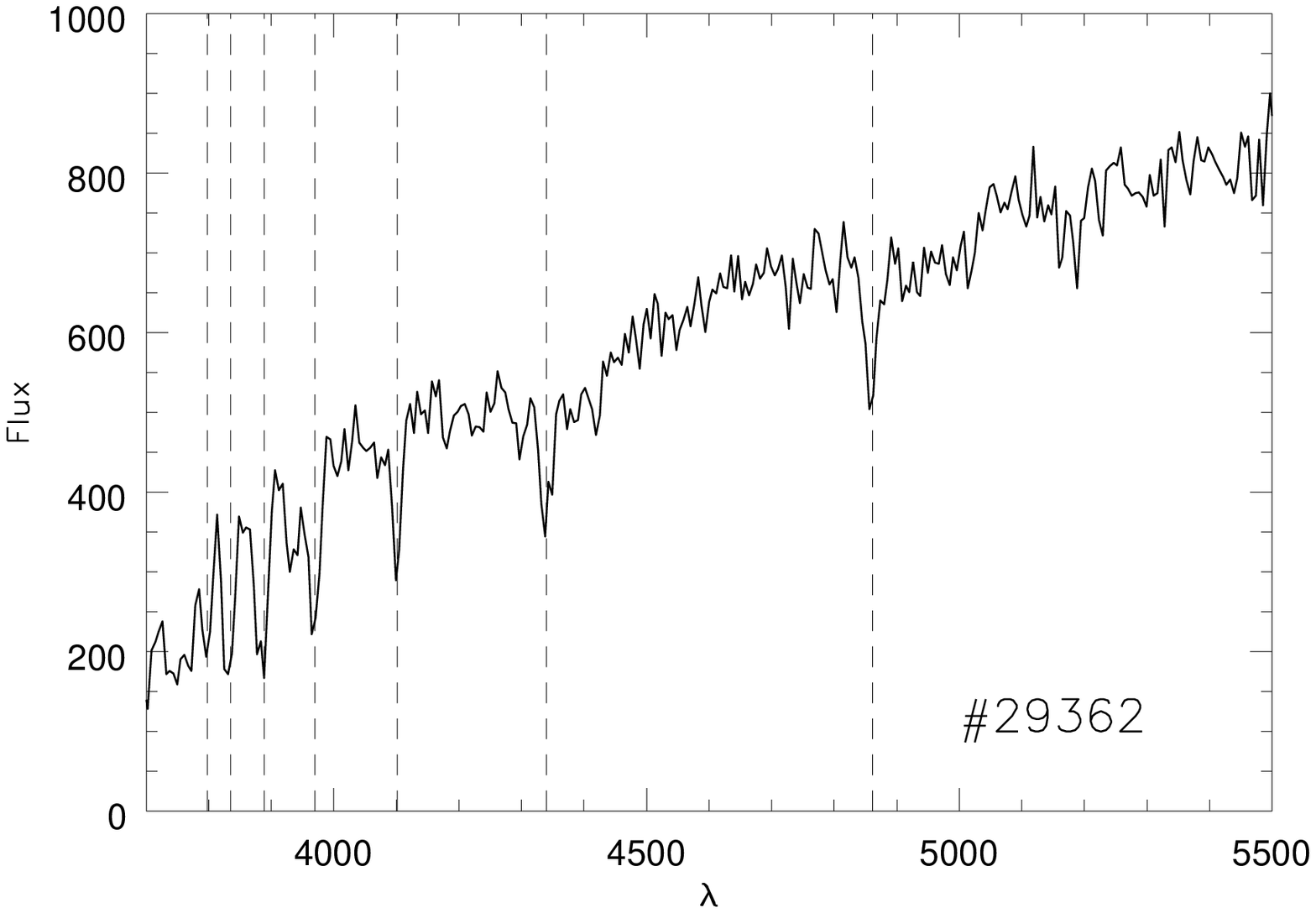}
\includegraphics[width=0.40\columnwidth,angle=0,clip]{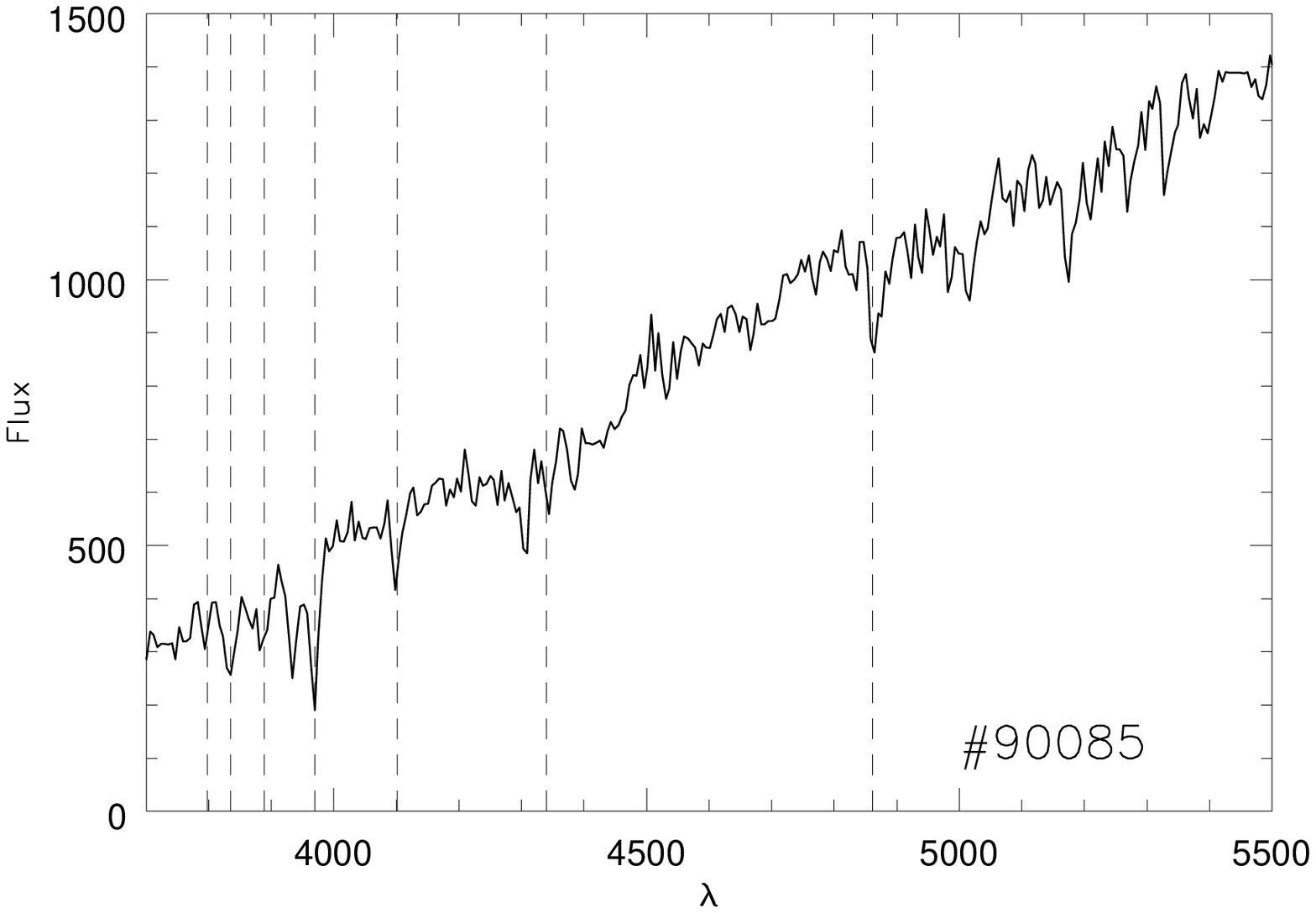}
\includegraphics[width=0.40\columnwidth,angle=0,clip]{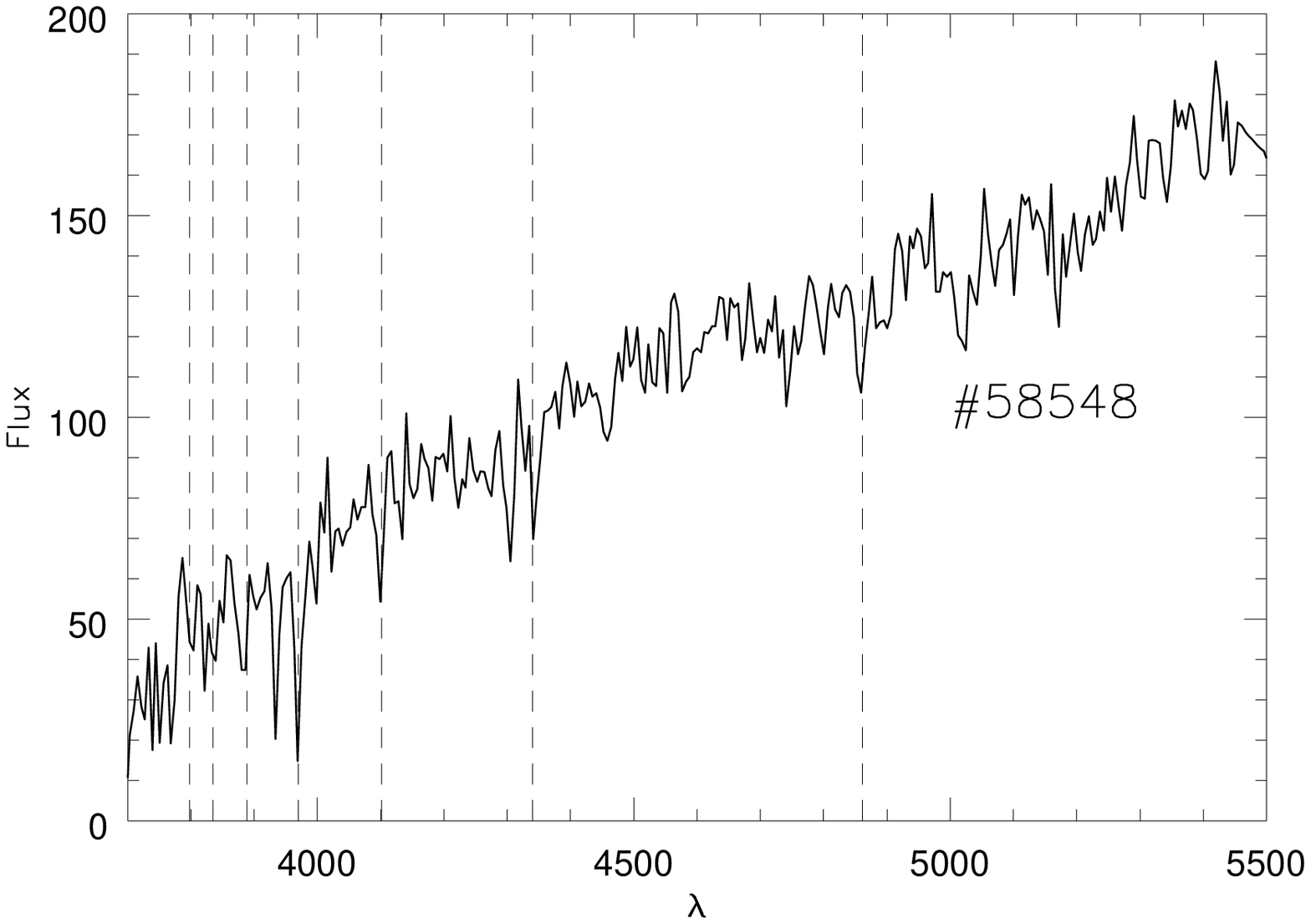}
\leftline{Fig.3 - Continued.}
\end{figure}

\begin{figure}
\centering
\includegraphics[width=0.40\columnwidth,angle=0,clip]{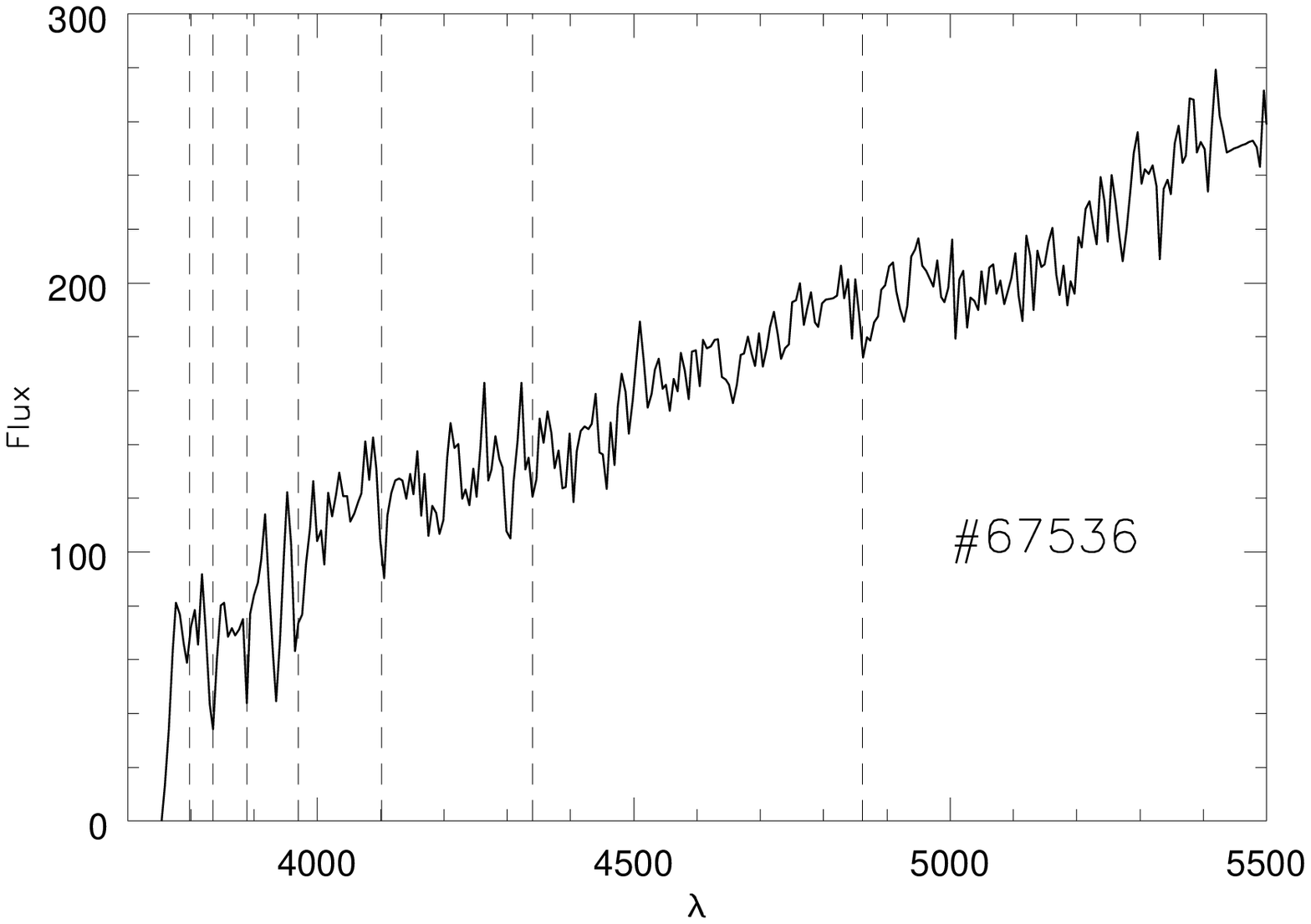}
\includegraphics[width=0.40\columnwidth,angle=0,clip]{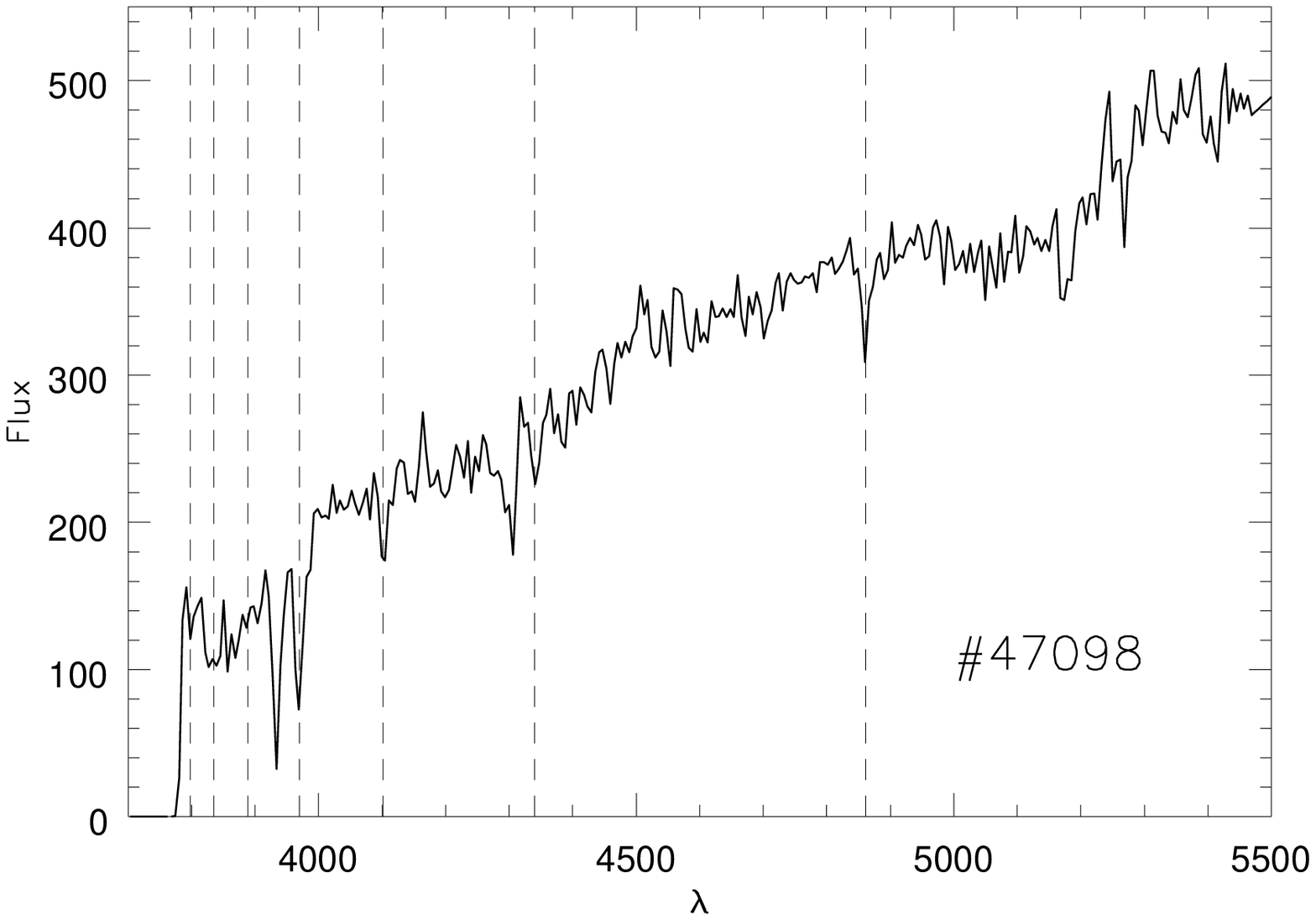}
\includegraphics[width=0.40\columnwidth,angle=0,clip]{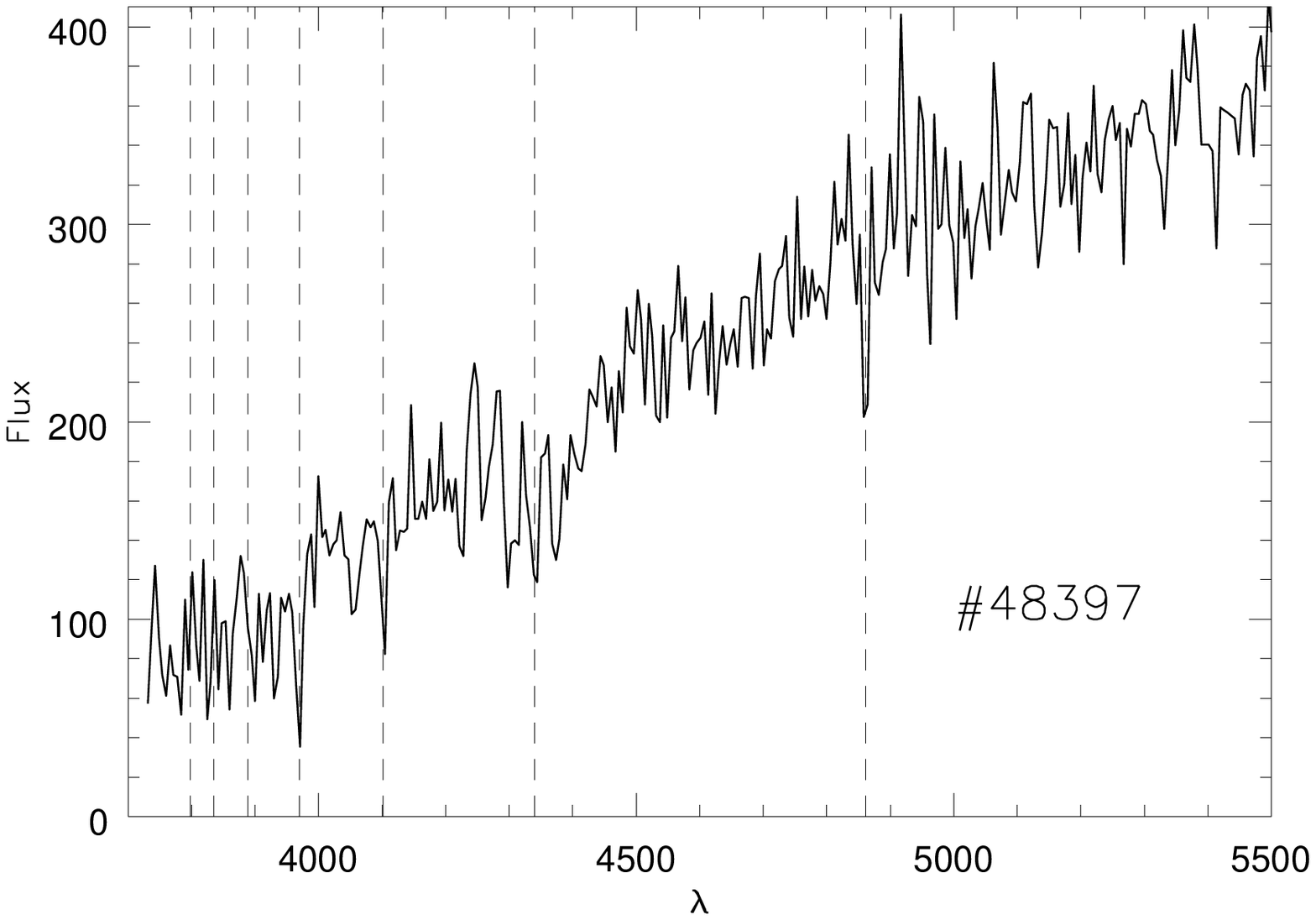}
\includegraphics[width=0.40\columnwidth,angle=0,clip]{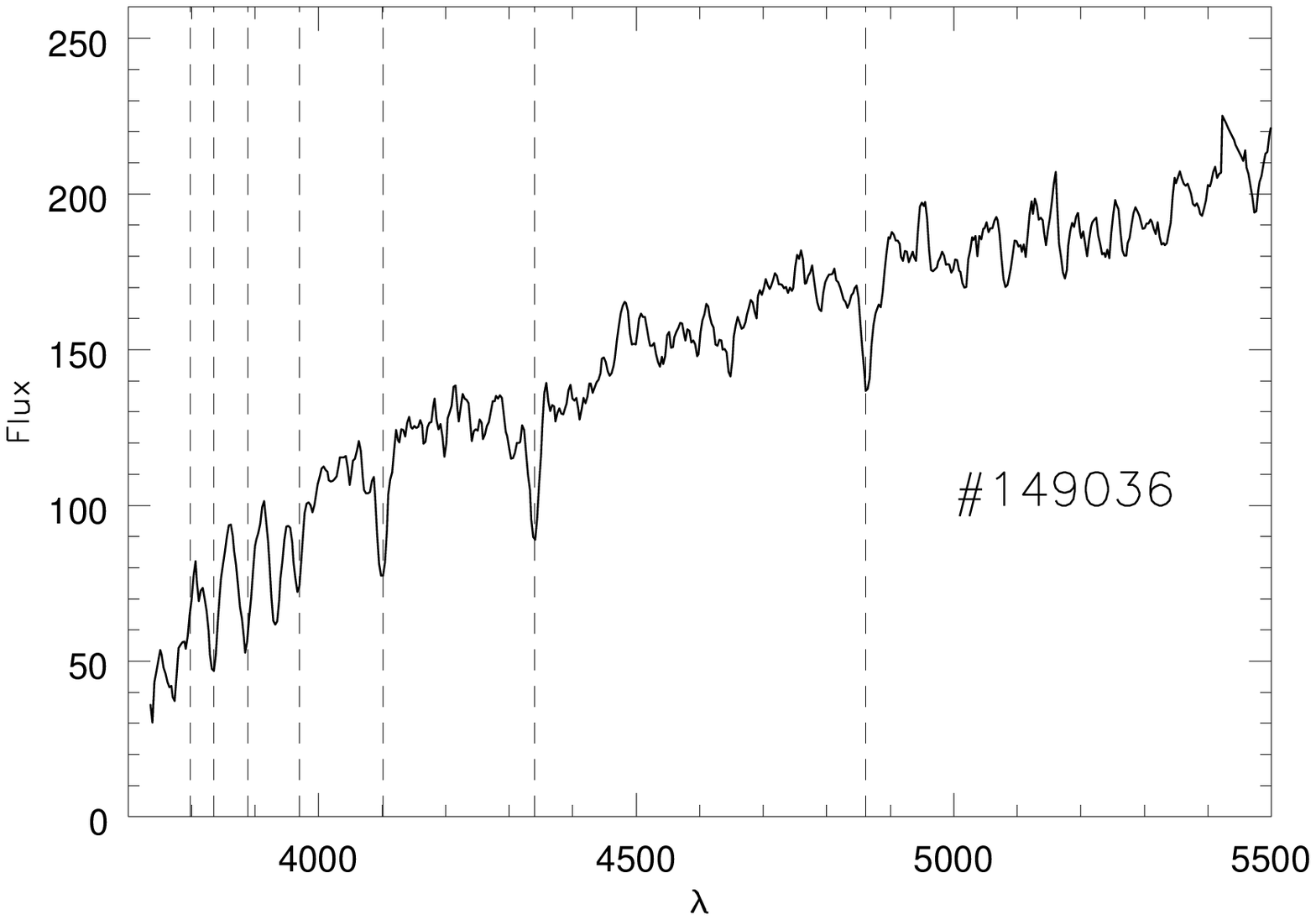}
\includegraphics[width=0.40\columnwidth,angle=0,clip]{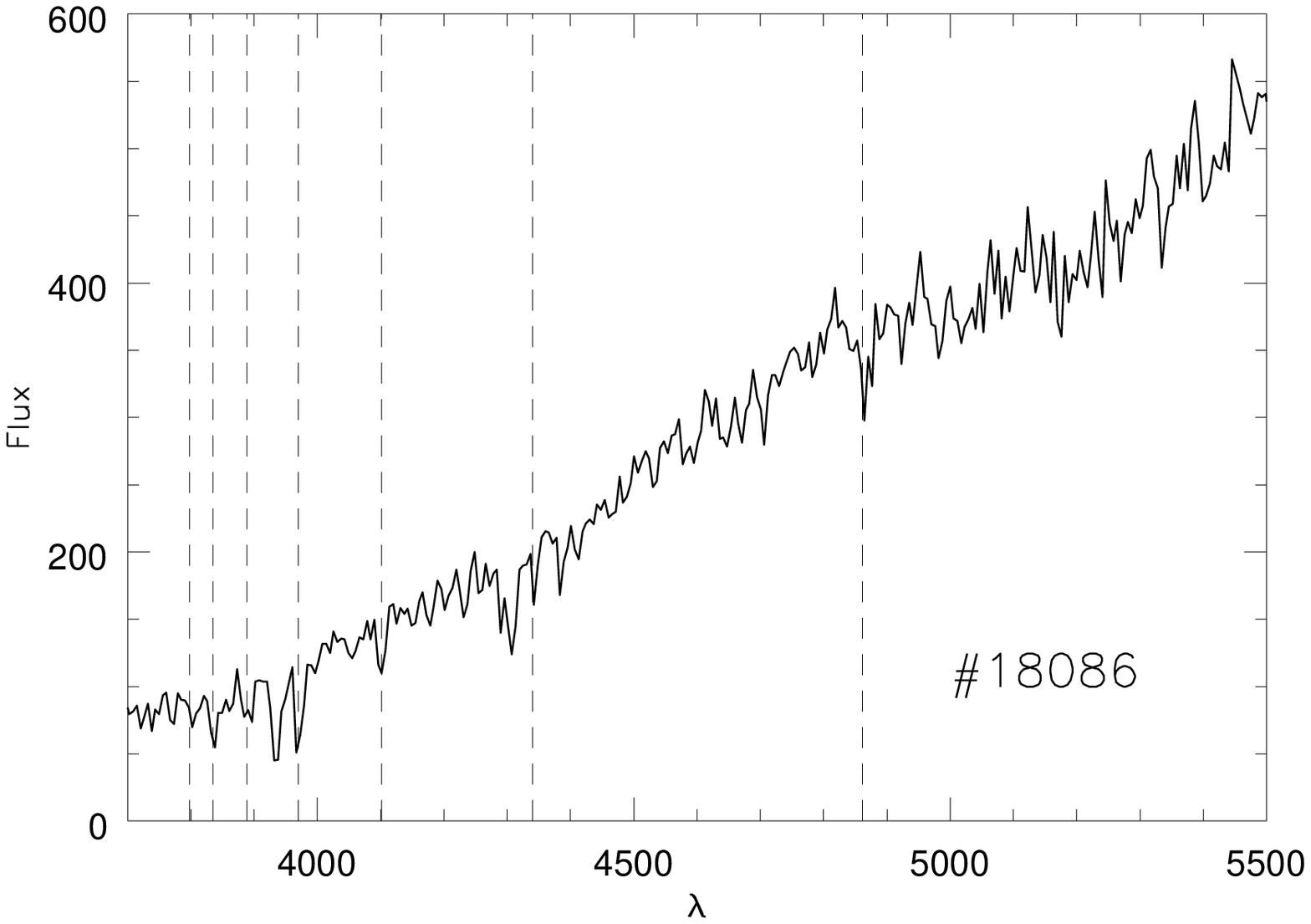}
\leftline{Fig.3 - Continued.}
\end{figure}


\begin{figure}
\epsscale{1.00}
\plotone{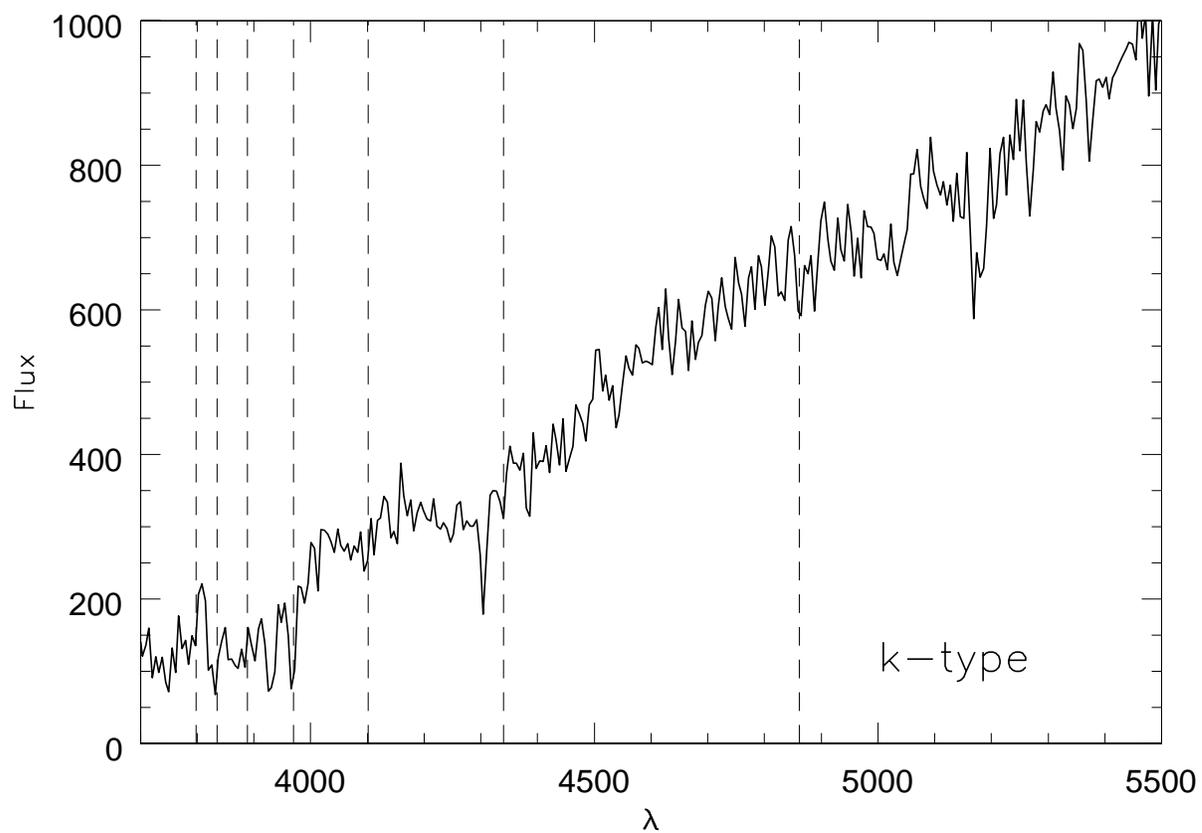}
\caption{Spectrum of a k-type galaxy with similar R magnitude
and S/N to the k+a spectra shown in the previous figure.
\label{fig3}}
\end{figure}

\clearpage 

\begin{figure}
\plotone{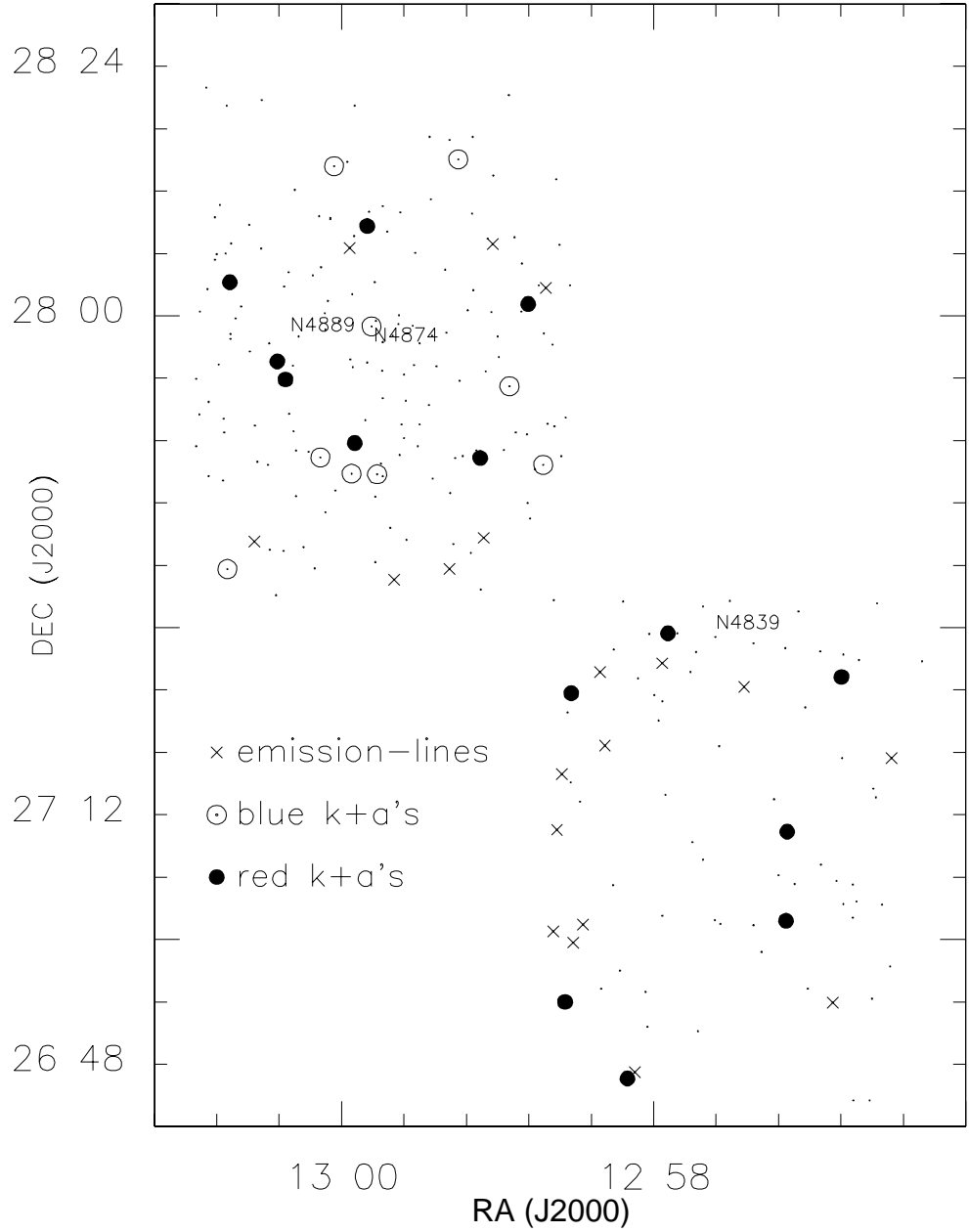}
\caption{Projected position on the sky of k (small dots),
red and blue k+a (large circles) and emission-line
(crosses) galaxies. The location of the three dominant galaxies
(N4874, N4889 and N4839) is labelled.
\label{fig3}}
\end{figure}

\clearpage 

\begin{figure}
\plotone{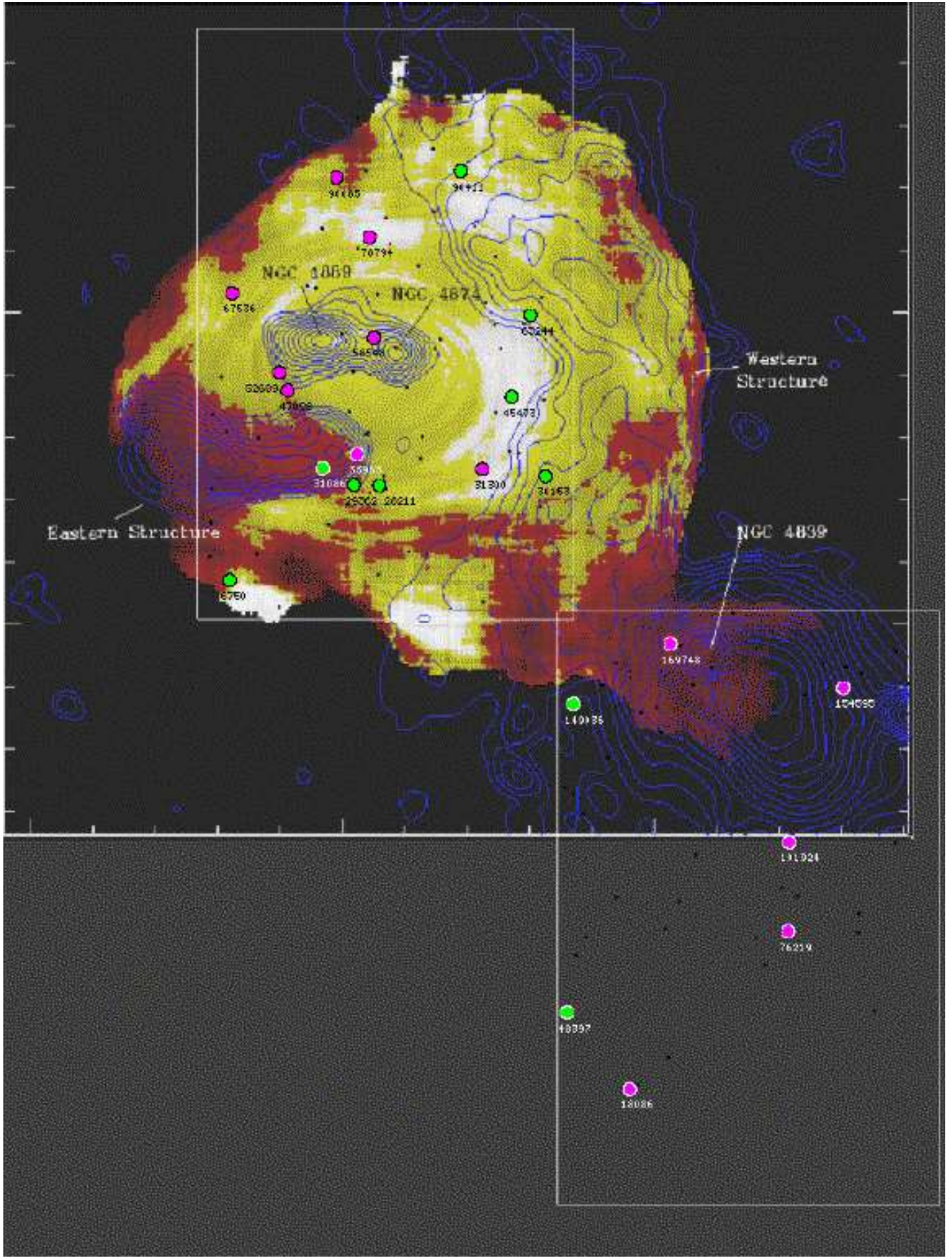}
\caption{Position of k+a galaxies with respect to X-ray substructure
and X-ray 
temperature map. Strong-lined k+a's with EW($\rm H\delta)>5$ \AA $\,$ 
are shown as green dots, while weaker k+a's are plotted as magenta dots.
Tiny black dots are dwarf Coma members with velocities $> 7200 \, \rm km \,
s^{-1}$.
X-ray residuals from \cite{neu03} are plotted as contours and clearly
identify two substructures (Western and Eastern substructures), in addition
to the NGC4839 peak in the South-West and the excess of emission
towards the two central galaxies (NGC4874 and NGC4889). 
The lowest contour and the step width
between two contours are each 5 $\sigma$. 
The hardness ratio image 
(2-5 keV/0.5-2keV, \cite{neu03})
is shown in color. Red regions correspond to temperatures
below 8 keV, yellow to $kT>8$ keV and white regions to $kT>10$ keV.
The rectangles show the limits of the two fields of our
photometric and spectroscopic survey (Coma1 towards the 
cluster center and Coma3 in the South-West). Each rectangle is 
about 1 by 1.5 Mpc.
\label{fig3}}
\end{figure}

\clearpage 

\begin{figure}
\plotone{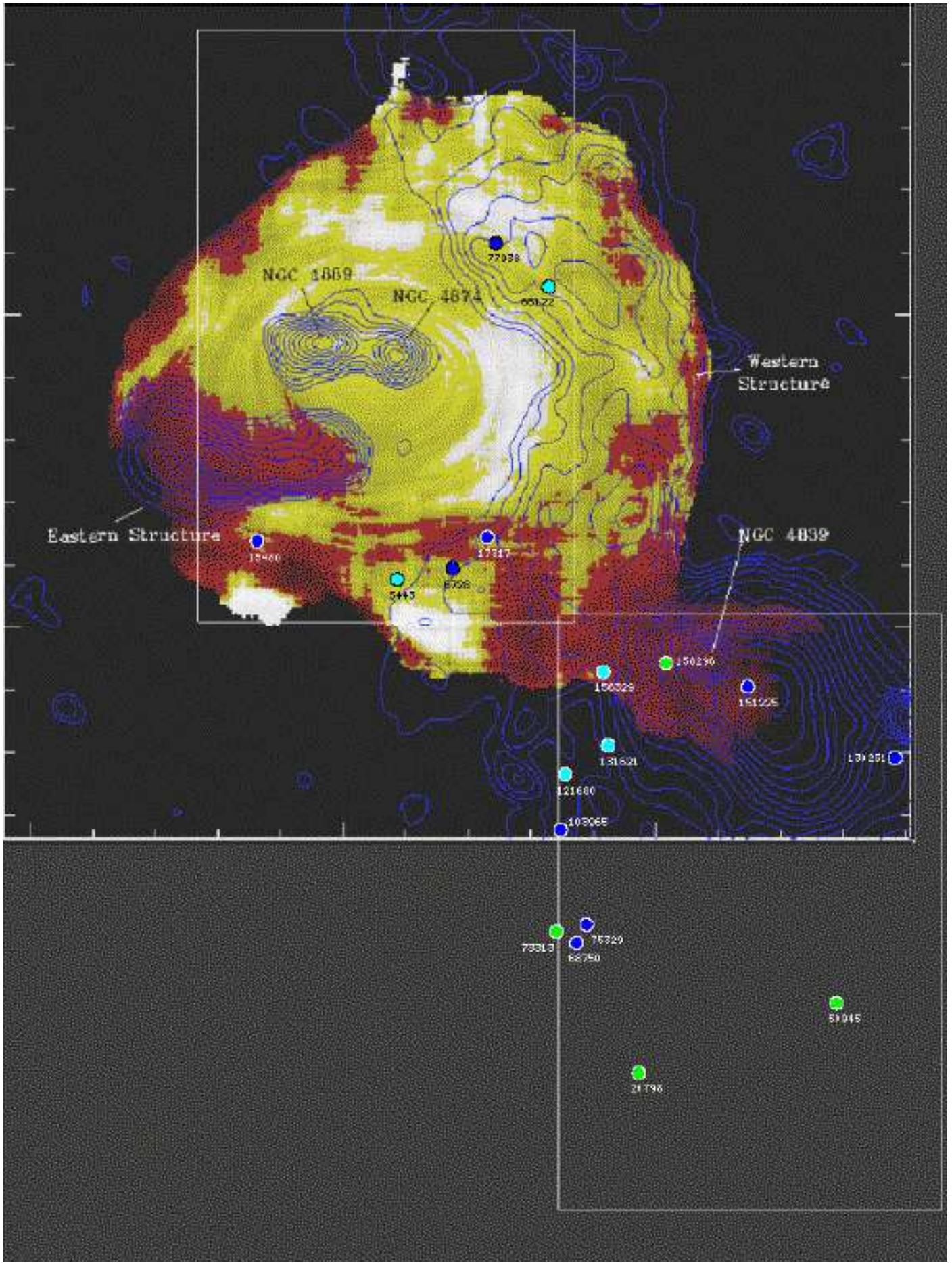}
\caption{Same as Fig.~6, but now showing the positions of emission-line 
galaxies instead of k+a's. Dark blue dots represent starburst galaxies, green
dots are quiescently star-forming galaxies (spiral-like) and light blue
dots are emission-line galaxies of unknown activity (either starbursts
or spiral-like).
\label{fig3}}
\end{figure}

\clearpage

\end{document}